# Towards a New Perturbation Theory of Molecular Liquids

Mehrdad Khanpour[a]

[a] *Department of Chemistry, Amol Branch, Islamic Azad University, Amol, Iran.*

**Abstract**

First steps towards developing a new perturbation theory for molecular liquids are taken. By choosing a new form of splitting the site-site potential functions between molecules, we will get a set of atomic fluids as the reference system with known structure and thermodynamics. The perturbative part of the potential function is then expanded up to two terms. The excess Helmholtz free energy of the system is then obtained for three computable contributions. The derivation shows that the excess Helmholtz free energy has nothing to do with intra-atomic potentials, all contributions come merely from inter-atomic potentials. Then it is applied to compute the thermodynamics of two systems; hard sphere chain and carbon dioxide molecular fluids. The results compared with the computer simulation data show that the theory works well at low densities.

E-mail: mtankhanpour@yahoo.com

# I. Introduction

Perturbation theories are probably the most successful in investigating the structure and thermodynamics of simple liquid matter. However, for real molecular liquids there is no such theory capable of finding their thermodynamic properties accurately except for model associating fluids. Unlike numerous perturbation theories for simple fluids [1-15], there is essentially one perturbation theory due to Wertheim for associating fluids with model simple potentials [16-20]. Although it can deliver very good results for model associating fluids, it cannot be applied for real complex liquids with more realistic potentials like those being employed in simulations. There is, of course, another theoretical method to perform such calculations, i.e., integral equation theories, usually appropriate for rigid molecules. Although they are constantly being developed, they have their own advantages and limitations [21-22] and are not our concern here. Because of this, most researchers resort to computer simulation methods to investigate the structure and thermodynamic properties of various forms of complex liquids. In order to model real systems as closely as possible in a simulation, one needs first to construct the molecules (out of atoms) and then define some reasonable potential functions (usually called force fields) acting between atoms belonging to the same or different molecules. In this way one can study almost every real system to obtain equilibrium, non-equilibrium, and structural properties usually very close to experimental ones. There is essentially no theory capable of working out real systems or even coming close to the simulation methods. We here want to take first steps towards developing a general perturbation theory for a pure molecular liquid composed of many different atoms interacting with 'real' potentials, i.e., force fields. We can achieve this goal by selecting a special group of potentials that enables us to define a set of atomic liquids with known properties as the appropriate reference system. Furthermore, this choice enables us to overcome the need for knowing more than two particle correlation functions when computing the perturbation terms. We will then compute explicitly the next two perturbation terms to obtain explicit expressions for the excess Helmholtz free energy of a completely general molecular fluid, those usually being worked out in simulations. The obtained Helmholtz free energy will then be used to obtain the thermodynamic properties of molecular liquids. All these will be performed in the next section devoted to derive the theory. As examples, the thermodynamic properties of a model system and a real system, i.e., hard sphere chain and carbon dioxide molecular fluids are investigated in later section. The strengths, weaknesses, limitations, and possible potentialities of the theory will be discussed in the last section.

## II. Theory

### 1. Some Required Preliminaries

As starting point we consider a single component molecular fluid composed of $N$ molecules each with $m$ (distinguishable) atoms contained in volume $V$ at temperature $T$. Similar to most computer simulation methods, we will invoke only site-site pair-potential functions. Let $u(|q_{i\mu} - q_{j\nu}|)$ represent the inter-potential function between atom $\mu$ in molecule $i$ located at $q_{i\mu}$ and atom $\nu$ in another molecule $j$ located at $q_{j\nu}$ and $u(|q_{k\gamma} - q_{k\delta}|)$ stands for intra-potential function between two bound atoms $\gamma$ (located at $q_{k\gamma}$) and $\delta$ (located at $q_{k\delta}$) within the same molecule $k$. We will then assume that these potentials (or force fields as they are usually called) are all pairwise. Hence, the net potential acting between atoms within and among the molecules reads: $U = \underbrace{\sum_{i<j}^{N}\sum_{\mu=1}^{m}\sum_{\nu=1}^{m} u(|q_{i\mu} - q_{j\nu}|)}_{inter-potentials} + \underbrace{\sum_{k=1}^{N}\sum_{\gamma,\delta} u(|q_{k\gamma} - q_{k\delta}|)}_{intra-potentials} = \sum_{\substack{i\leq j \\ \mu,\nu}}^{N} u(|q_{i\mu} - q_{j\nu}|)$ where in the last expression equality sign is employed to merge both inter and intra potential contributions. We must always bear in mind that in case of intra-molecular potentials the atomic indexes are representing only bound atoms. Note also that we do not impose any other restriction on potential functions or force fields acting on different atoms. The partition function of this system in the canonical ensemble (ignoring all ideal contributions that are assumed to be known) reads:

$$Q = \frac{1}{V^{Nm}} \int e^{-\beta \sum_{\substack{i \leq j \\ \mu,\nu}}^{N} u(|q_{i\mu} - q_{j\nu}|)} \underbrace{(dq_{11} dq_{21} \ldots dq_{N1})}_{dq_1^N} \underbrace{(dq_{12} dq_{22} \ldots dq_{N2})}_{dq_2^N} \ldots \underbrace{(dq_{1m} dq_{2m} \ldots dq_{Nm})}_{dq_m^N} \quad (1).$$

Here $\beta$ is equal to $1/k_B T$ where $k_B$ is the Boltzmann constant and $q_{i\mu}(q_{j\nu})$ stands for the position of atom $\mu$ ($\nu$) within the molecule $i$ ($j$). Note that we have put aside all ideal contributions (an $N!$ factor, all momenta contributions, etc.) in Eq. (1). Hence all thermodynamic functions due to this partition function would be excess with respect to ideal contributions. That is we are to find the excess Helmholtz free energy and hence other excess thermodynamic properties.

## 2. Reference System and Perturbation Theory

In every perturbation theory we must have a reference system with known properties. Here, we will choose the potentials between the same atoms in different molecules as the reference potential. That is, for reference system we choose $U_0 = \sum_{\alpha=1}^{m} U_\alpha = \sum_{\alpha=1}^{m} \sum_{i<j}^{N} u(|q_{i\alpha} - q_{j\alpha}|)$ which contains the direct terms (including $m$ potential functions) between two molecules. Clearly $U_\alpha = \sum_{i<j}^{N} u(|q_{i\alpha} - q_{j\alpha}|)$ shows the potential between atoms in an atomic fluid containing exclusively of $\alpha$ atoms. The cross terms (incorporating $m(m-1)$ inter-potential functions) and also intra-potentials are taken as perturbation terms. That is, we will split the whole potential such that we have:

$$U = \underbrace{\sum_{\alpha=1}^{m} \sum_{i<j}^{N} u(|q_{i\alpha} - q_{j\alpha}|)}_{U_0} + \sum_{i<j}^{N} \sum_{\mu=1}^{m} \sum_{\nu(\neq\mu)}^{m} u(|q_{i\mu} - q_{j\nu}|) + \sum_{k=1}^{N} \sum_{\gamma,\delta} u(|q_{k\gamma} - q_{k\delta}|) \quad (2)$$

The most important reason for this choice is that one can immediately calculate the partition function given by Eq. (1) by use of $U_0$ as the mere potential acting between atoms in molecules. That is, the partition function of the reference system will be:

$$Q_0 = \frac{1}{V^{Nm}} \int e^{-\beta \sum_{\alpha=1}^{m} U_\alpha} dq_1^N ... dq_m^N = \left( \frac{1}{V^N} \int e^{-\beta U_1} dq_1^N \right) ... \left( \frac{1}{V^N} \int e^{-\beta U_m} dq_m^N \right) = Q_{01} ... Q_{0m} \qquad (3)$$

in which $Q_{0\alpha} = \frac{1}{V^N} \int e^{-\beta \sum_{i<j}^{N} u(|q_{i\alpha}-q_{j\alpha}|)} dq_\alpha^N = \frac{1}{V^N} \int e^{-\beta U_\alpha} dq_\alpha^N$ is the partition function of atomic fluid containing only $\alpha$ atoms. So the reference system is a set of $m$ distinct atomic fluids with known structure and thermodynamics. Another advantage of our choice comes from the following observation on one and two atom density functions. The one particle density function is defined as:

$$\rho^{(1)}(q_{1\mu}) = \frac{N}{V^{Nm}} \int \frac{e^{-\beta \sum_{i<j,\mu,\nu}^{N} u(|q_{i\mu}-q_{j\nu}|)}}{Q} dq_1^N ... dq_\mu^{N-1} ... dq_m^N \qquad (4)$$

For reference system it transforms into:

$$\rho^{(1)}(q_{1\mu}) = \frac{N}{V^{Nm}} \int \frac{e^{-\beta \sum_{\alpha=1}^{m} U_\alpha}}{Q_0} dq_1^N ... dq_\mu^{N-1} ... dq_m^N = \left( \frac{1}{V^N} \int \frac{e^{-\beta U_1}}{Q_{01}} dq_1^N \right) ... \left( \frac{N}{V^N} \int \frac{e^{-\beta U_\mu}}{Q_{0\mu}} dq_\mu^{N-1} \right)$$
$$... \left( \frac{1}{V^N} \int \frac{e^{-\beta U_m}}{Q_{0m}} dq_m^N \right) = \frac{N}{V^N} \int \frac{e^{-\beta U_\mu}}{Q_{0\mu}} dq_\mu^{N-1} \qquad (5).$$

That is, the one particle density function of reference system equals to the one particle density function of its corresponding atomic fluid. Contrary to one particle density function, there are essentially two pair correlation functions related to the two atoms that are singled out from two separate molecules (we will ignore these for atoms belonging to one molecule). For two different atoms (located on two different molecules) it is defined as:

$$\rho^{(2)}(q_{1\mu}, q_{2\nu}) = \frac{N^2}{V^{Nm}} \int \frac{e^{-\beta \sum_{i<j,\mu,\nu}^{N} u(|q_{i\mu}-q_{j\nu}|)}}{Q} dq_1^N ... dq_\mu^{N-1} ... dq_\nu^{N-1} ... dq_m^N \qquad (6a)$$

And when the two atoms (located on two different molecules) are of the same kind:

$$\rho^{(2)}(q_{1\mu}, q_{2\mu}) = \frac{N(N-1)}{V^{Nm}} \int \frac{e^{-\beta \sum_{\substack{i \leq j \\ \mu,\nu}}^{N} u(|q_{i\mu} - q_{j\nu}|)}}{Q} dq_1^N \ldots dq_\mu^{N-2} \ldots dq_m^N \qquad (6b).$$

For the chosen reference system these quantities are reduced to:

$$\rho^{(2)}(q_{1\mu}, q_{2\nu}) = \frac{N^2}{V^{Nm}} \int \frac{e^{-\beta \sum_{\alpha=1}^{m} U_\alpha}}{Q} dq_1^N \ldots dq_\mu^{N-1} \ldots dq_\nu^{N-1} \ldots dq_m^N = \left(\frac{1}{V^N} \int \frac{e^{-\beta U_1}}{Q_{01}} dq_1^N\right)$$
$$\left(\frac{N}{V^N} \int \frac{e^{-\beta U_\mu}}{Q_{0\mu}} dq_\mu^{N-1}\right) \ldots \left(\frac{N}{V^N} \int \frac{e^{-\beta U_\nu}}{Q_{0\nu}} dq_\nu^{N-1}\right) \ldots \left(\frac{1}{V^N} \int \frac{e^{-\beta U_m}}{Q_{0m}} dq_m^N\right) = \rho^{(1)}(q_{1\mu})\rho^{(1)}(q_{2\nu}) \qquad (7)$$

and more important:

$$\rho^{(2)}(q_{1\mu}, q_{2\mu}) = \frac{N(N-1)}{V^{Nm}} \int \frac{e^{-\beta \sum_{\alpha=1}^{m} U_\alpha}}{Q_0} dq_1^N \ldots dq_\mu^{N-2} \ldots dq_m^N = \left(\frac{1}{V^N} \int \frac{e^{-\beta U_1}}{Q_{01}} dq_1^N\right) \ldots$$
$$\left(\frac{N(N-1)}{V^N} \int \frac{e^{-\beta U_\mu}}{Q_{0\mu}} dq_\mu^{N-2}\right) \ldots \left(\frac{1}{V^N} \int \frac{e^{-\beta U_m}}{Q_{0m}} dq_m^N\right) = \rho^{(2)}(q_{1\mu}, q_{2\mu}) = \rho^{(1)}(q_{1\mu})\rho^{(1)}(q_{2\mu})g(q_{1\mu}, q_{2\mu}) \qquad (8)$$

where in the last equality we have utilized the definition of the pair correlation function of atomic fluids. Furthermore, we will write $g(q_{1\mu}, q_{2\mu}) = g_\mu(|q_{1\mu} - q_{2\mu}|)$ for our system because of the fact that the pair potential between the same atoms in different molecules depends only on their distance and hence for their pair correlation functions. Therefore, two-particle density functions of a molecular fluid are reduced to the already known radial distribution functions of $m$ atomic fluids comprising our reference system. Since both thermodynamics and structure functions are known for our reference system we should be able to use them as our main means in a perturbation theory. In what follows we will show that we can proceed and compute up to two terms in high temperature expansion form of perturbation theory. To be complete we first present the main elements of high temperature perturbation theory. Let the partition function of a system be written as:

$$Q(\xi) = Q_0 + \xi Q_1 + \frac{\xi^2}{2} Q_2 \tag{9}$$

Where $\xi$ represent the coupling parameter. When $\xi = 1$ it represents the partition function of the whole system whereas at $\xi = 0$ it returns to the reference system. The Helmholtz free energy of the system will then become $A(\xi) = -\frac{1}{\beta} \ln Q(\xi) = -\frac{1}{\beta} \ln \left( Q_0 + \xi Q_1 + \frac{\xi^2}{2} Q_2 \right)$. Expanding of $A(\xi)$ in terms of $\xi$ up to two terms gives $A(\xi) = A_0 - \frac{\xi}{\beta} \left( \frac{Q_1}{Q_0} \right) - \frac{\xi^2}{2\beta} \left( \frac{Q_2}{Q_0} - \left( \frac{Q_1}{Q_0} \right)^2 \right)$ where $A_0 = -\frac{1}{\beta} \ln Q_0$. Let us express it in terms of the reduced Helmholtz free energy per particle (or molecule) defined as $a = \frac{\beta A}{N}$ and then put $\xi = 1$ to obtain:

$$a = a_0 - \frac{1}{N}\left(\frac{Q_1}{Q_0}\right) - \frac{1}{2N}\left( \frac{Q_2}{Q_0} - \left(\frac{Q_1}{Q_0}\right)^2 \right) \tag{10}$$

Since the Helmholtz free energy of all atomic liquids are extensive and proportional to $N$, we have then ignored it in the first reference term. Furthermore, as the above equation shows the perturbation terms should be proportional to $N$. The first and second perturbation terms are thus:

$$a_1 = \frac{1}{N}\frac{Q_1}{Q_0} \quad \text{and} \quad a_2 = \frac{1}{N}\left( \frac{Q_2}{Q_0} - \left(\frac{Q_1}{Q_0}\right)^2 \right) \tag{11}$$

So, the reduced Helmholtz free energy of the whole system of a molecular fluid will be (ignoring $N$):

$$a = a_0 - a_1 - \frac{a_2}{2} \tag{12}$$

Based on what has been done so far, one can guess that all three contributions are computable. In the following we will see that this is true. One should also note that only terms proportional to $N$ must be maintained to remove $N$ in Eq. (10). All other terms will vanish in the thermodynamic limit. As said before, the first term corresponding to the reduced free energy of the reference system is simply given by $Q_0 = Q_{01} \cdots Q_{0m}$, and hence we will simply find $a_0 = \sum_{\alpha=1}^{m} a_{0\alpha}$, where $a_{0\alpha}$ represent the reduced free energy of the atomic fluid composed of $\alpha$ atoms with potential $U_\alpha$. In order to proceed further we must know explicitly $Q_1$ and $Q_2$. The partition function of a molecular fluid reads:

$$Q = \frac{1}{V^{Nm}} \int e^{-\beta \sum_{\substack{i \leq j \\ \mu,\nu}}^{N} u(|q_{i\mu}-q_{j\nu}|)} dq_1^N \ldots dq_m^N = \frac{1}{V^{Nm}} \int e^{-\beta \sum_{\alpha=1}^{m} U_\alpha} e^{-\beta \sum_{\substack{i \leq j \\ \mu \neq \nu}}^{N} u(|q_{i\mu}-q_{j\nu}|)} dq_1^N \ldots dq_m^N .$$ In terms of Mayer functions

(defined by $f_{\mu\nu}(q) = e^{-\beta u_{\mu\nu}(q)} - 1$) we have:

$$Q = \frac{1}{V^{Nm}} \int e^{-\beta \sum_{\alpha=1}^{m} U_\alpha} e^{-\beta \sum_{\substack{i \leq j \\ \mu \neq \nu}}^{N} u(|q_{i\mu}-q_{j\nu}|)} dq_1^N \ldots dq_m^N = \frac{1}{V^{Nm}} \int e^{-\beta \sum_{\alpha=1}^{m} U_\alpha} \prod_{\substack{i \leq j \\ \mu \neq \nu}} \left(1 + f_{\mu\nu}(|q_{i\mu} - q_{j\nu}|)\right) dq_1^N \ldots dq_m^N$$

In order to transform it to the form of Eq. (9), we must introduce a coupling factor such that at $\xi = 1$ it represents the real system whereas at $\xi = 0$ reduces to reference system. We thus generalize the partition function in the following form and expand it up to second term:

$$Q(\xi) = \frac{1}{V^{Nm}} \int e^{-\beta \sum_{\alpha=1}^{m} U_\alpha} \prod_{\substack{i \leq j \\ \mu \neq \nu}} \left(1 + \xi f_{\mu\nu}(|q_{i\mu} - q_{j\nu}|)\right) dq_1^N \ldots dq_m^N =$$

$$\frac{1}{V^{Nm}} \int e^{-\beta \sum_{\alpha=1}^{m} U_\alpha} \left\{ 1 + \xi \sum_{\substack{i \leq j \\ \mu \neq \nu}} f_{\mu\nu}(|q_{i\mu} - q_{j\nu}|) + \frac{\xi^2}{2} \sum_{\substack{i \leq j \\ \mu \neq \nu}} \sum_{\substack{i \leq j \\ \mu' \neq \nu'}} f_{\mu\nu}(|q_{i\mu} - q_{j\nu}|) f_{\mu'\nu'}(|q_{i\mu'} - q_{j\nu'}|) + \ldots \right\} dq_1^N \ldots dq_m^N \quad (13).$$

We thus find:

$$Q_1 = \frac{1}{V^{Nm}} \sum_{\substack{i \leq j \\ \mu \neq \nu}} \int e^{-\beta \sum_{\alpha=1}^{m} U_\alpha} f_{\mu\nu}(|q_{i\mu} - q_{j\nu}|) dq_1^N ... dq_m^N \qquad (14)$$

and

$$Q_2 = \frac{1}{V^{Nm}} \sum_{\substack{i \leq j \\ \mu \neq \nu}} \sum_{\substack{i' \leq j' \\ \mu' \neq \nu'}} \int e^{-\beta \sum_{\alpha=1}^{m} U_\alpha} f_{\mu\nu}(|q_{i\mu} - q_{j\nu}|) f_{\mu'\nu'}(|q_{i'\mu'} - q_{j'\nu'}|) dq_1^N ... dq_m^N \qquad (15).$$

Care should be paid in treating the inter-atomic and intra-atomic potential contributions. We will then separate them as follows:

$$\sum_{\substack{i \leq j \\ \mu \neq \nu}} f_{\mu\nu}(|q_{i\mu} - q_{j\nu}|) = \sum_{\substack{i < j \\ \mu \neq \nu}} f_{\mu\nu}(|q_{i\mu} - q_{j\nu}|) + \sum_{\substack{k \\ \gamma \neq \delta}} f_{\gamma\delta}(|q_{k\gamma} - q_{k\delta}|)$$

This gives:

$$\sum_{\substack{i \leq j \\ \mu \neq \nu}} \sum_{\substack{i' \leq j' \\ \mu' \neq \nu'}} f_{\mu\nu}(|q_{i\mu} - q_{j\nu}|) f_{\mu'\nu'}(|q_{i'\mu'} - q_{j'\nu'}|) = \left( \sum_{\substack{i < j \\ \mu \neq \nu}} f_{\mu\nu}(|q_{i\mu} - q_{j\nu}|) + \sum_{\substack{k \\ \gamma \neq \delta}} f_{\gamma\delta}(|q_{k\gamma} - q_{k\delta}|) \right) \left( \sum_{\substack{i' < j' \\ \mu' \neq \nu'}} f_{\mu'\nu'}(|q_{i'\mu'} - q_{j'\nu'}|) + \sum_{\substack{k' \\ \gamma' \neq \delta'}} f_{\gamma'\delta'}(|q_{k'\gamma'} - q_{k'\delta'}|) \right)$$

$$= \left( \sum_{\substack{i < j \\ \mu \neq \nu}} \sum_{\substack{i' < j' \\ \mu' \neq \nu'}} f_{\mu\nu}(|q_{i\mu} - q_{j\nu}|) f_{\mu'\nu'}(|q_{i'\mu'} - q_{j'\nu'}|) \right) + 2 \left( \sum_{\substack{i < j \\ \mu \neq \nu}} \sum_{\substack{k \\ \gamma \neq \delta}} f_{\mu\nu}(|q_{i\mu} - q_{j\nu}|) f_{\gamma\delta}(|q_{k\gamma} - q_{k\delta}|) \right) + \left( \sum_{\substack{k \\ \gamma \neq \delta}} \sum_{\substack{k' \\ \gamma' \neq \delta'}} f_{\gamma\delta}(|q_{k\gamma} - q_{k\delta}|) f_{\gamma'\delta'}(|q_{k'\gamma'} - q_{k'\delta'}|) \right)$$

Thus the first perturbation term has two contributions:

$$Q_1 = Q_{11} + Q_{12}$$
$$= \frac{1}{V^{Nm}} \sum_{\substack{i < j \\ \mu \neq \nu}} \int e^{-\beta \sum_{\alpha=1}^{m} U_\alpha} f_{\mu\nu}(|q_{i\mu} - q_{j\nu}|) dq_1^N ... dq_m^N + \frac{1}{V^{Nm}} \sum_{\substack{k \\ \gamma \neq \delta}} \int e^{-\beta \sum_{\alpha=1}^{m} U_\alpha} f_{\gamma\delta}(|q_{k\gamma} - q_{k\delta}|) dq_1^N ... dq_m^N \qquad (16).$$

The second perturbation term, however, has three contributions, i.e., $Q_2 = Q_{21} + 2Q_{22} + Q_{23}$ where:

$$Q_{21} = \frac{1}{V^{Nm}} \sum_{\substack{i<j \\ \mu \neq \nu}} \sum_{\substack{i'<j' \\ \mu' \neq \nu'}} \int e^{-\beta \sum_{\alpha=1}^{m} U_\alpha} f_{\mu\nu}(|q_{i\mu} - q_{j\nu}|) f_{\mu'\nu'}(|q_{i'\mu'} - q_{j'\nu'}|) dq_1^N ... dq_m^N \qquad (17)$$

$$Q_{22} = \frac{1}{V^{Nm}} \sum_{\substack{i<j \\ \mu \neq \nu}} \sum_{\substack{k \\ \gamma \neq \delta}} \int e^{-\beta \sum_{\alpha=1}^{m} U_\alpha} f_{\mu\nu}(|q_{i\mu} - q_{j\nu}|) f_{\gamma\delta}(|q_{k\gamma} - q_{k\delta}|) dq_1^N ... dq_m^N \qquad (18)$$

and

$$Q_{23} = \frac{1}{V^{Nm}} \sum_{\substack{k \\ \gamma \neq \delta}} \sum_{\substack{k' \\ \gamma' \neq \delta'}} \int e^{-\beta \sum_{\alpha=1}^{m} U_\alpha} f_{\gamma\delta}(|q_{k\gamma} - q_{k\delta}|) f_{\gamma'\delta'}(|q_{k'\gamma'} - q_{k'\delta'}|) dq_1^N ... dq_m^N \qquad (19).$$

Corresponding to these expressions the first and second perturbation terms are expressed as follows:

$$a_1 = \frac{Q_1}{Q_0} = \frac{Q_{11} + Q_{12}}{Q_0} = a_{11} + a_{12} =$$

$$\frac{1}{V^{Nm}} \sum_{\substack{i<j \\ \mu \neq \nu}} \int \frac{e^{-\beta \sum_{\alpha=1}^{m} U_\alpha}}{Q_0} f_{\mu\nu}(|q_{i\mu} - q_{j\nu}|) dq_1^N ... dq_m^N + \frac{1}{V^{Nm}} \sum_{\substack{k \\ \gamma \neq \delta}} \int \frac{e^{-\beta \sum_{\alpha=1}^{m} U_\alpha}}{Q_0} f_{\gamma\delta}(|q_{k\gamma} - q_{k\delta}|) dq_1^N ... dq_m^N \qquad (20)$$

and

$$a_2 = \frac{Q_2}{Q_0} - \left(\frac{Q_1}{Q_0}\right)^2 = \frac{Q_{21} + 2Q_{22} + Q_{23}}{Q_0} - \left(\frac{Q_{11} + Q_{12}}{Q_0}\right)^2 =$$

$$= \underbrace{\left(\frac{Q_{21}}{Q_0} - \left(\frac{Q_{11}}{Q_0}\right)^2\right)}_{a_{21}} + 2\underbrace{\left(\frac{Q_{22}}{Q_0} - \left(\frac{Q_{11}Q_{12}}{Q_0^2}\right)\right)}_{a_{22}} + \underbrace{\left(\frac{Q_{23}}{Q_0} - \left(\frac{Q_{12}}{Q_0}\right)^2\right)}_{a_{23}} = a_{21} + 2a_{22} + a_{23} \qquad (21)$$

in combination with Eqs. (17), (18), and (19).

Now we are ready to calculate perturbation terms.

**3. Calculation of the First Perturbation Contributions to the Reduced Free Energy of the System**

The first perturbation term has two contributions $a_{11}$ and $a_{12}$. The $a_{11}$ term given by:

$$a_{11} = \frac{Q_1}{Q_0} = \frac{1}{V^{Nm}} \sum_{\substack{i<j \\ \mu \neq \nu}} \int \frac{e^{-\beta \sum_{\alpha=1}^{m} U_\alpha}}{Q_{01} \cdots Q_{0m}} f_{\mu\nu}(|q_{i\mu} - q_{j\nu}|) dq_1^N \cdots dq_m^N$$

can be calculated as follows:

$$a_{11} = \frac{N(N-1)}{2} \sum_{\mu \neq \nu} \int \frac{e^{-\beta U_1} \cdots e^{-\beta U_\mu} \cdots e^{-\beta U_\nu} \cdots e^{-\beta U_m}}{Q_{01} \cdots Q_{0\mu} \cdots Q_{0\nu} \cdots Q_{0m}} f_{\mu\nu}(|q_{1\mu} - q_{2\nu}|) dq_1^N \cdots dq_m^N$$

$$= \frac{1}{2} \sum_{\mu \neq \nu} \int dq_{1\mu} dq_{2\nu} f_{\mu\nu}(|q_{1\mu} - q_{2\nu}|) \underbrace{\left( \frac{1}{V^N} \int \frac{e^{-\beta U_1}}{Q_{01}} dq_1^N \right)}_{1} \cdots \underbrace{\left( \frac{N}{V^N} \int \frac{e^{-\beta U_\mu}}{Q_{0\mu}} dq_\mu^{N-1} \right)}_{\rho^{(1)}(q_{1\mu})} \cdots \underbrace{\left( \frac{N-1}{V^N} \int \frac{e^{-\beta U_\nu}}{Q_{0\nu}} dq_\nu^{N-1} \right)}_{\rho^{(1)}(q_{2\nu})} \cdots \underbrace{\left( \frac{1}{V^N} \int \frac{e^{-\beta U_m}}{Q_{0m}} dq_m^N \right)}_{1}$$

Hence:

$$a_{11} = \frac{1}{2} \sum_{\mu \neq \nu} \int dq_{1\mu} dq_{2\nu} f_{\mu\nu}(|q_{1\mu} - q_{2\nu}|) \rho^{(1)}(q_{1\mu}) \rho^{(1)}(q_{2\nu}) \tag{22}$$

For fluid systems $\rho^{(1)}(q_{1\mu}) = \rho^{(1)}(q_{2\nu}) = \rho$ where $\rho = \frac{N}{V}$ is the number density of the molecular fluid in a container with volume $V$. Thus we have:

$$a_{11} = \frac{N\rho}{2} \sum_{\mu \neq \nu} \int dq' f_{\mu\nu}(|q'|) \tag{23}.$$

The second contribution is calculated as follows:

$$a_{12} = \frac{1}{V^{Nm}} \sum_{\substack{k \\ \gamma \neq \delta}} \int \frac{e^{-\beta \sum_{\alpha=1}^{m} U_\alpha}}{Q_0} f_{\gamma\delta}(|q_{k\gamma} - q_{k\delta}|) dq_1^N \cdots dq_m^N =$$

$$= N \sum_{\gamma \neq \delta} \int \frac{e^{-\beta U_1} \cdots e^{-\beta U_\gamma} \cdots e^{-\beta U_\delta} \cdots e^{-\beta U_m}}{Q_{01} \cdots Q_{0\gamma} \cdots Q_{0\delta} \cdots Q_{0m}} f_{\gamma\delta}(|q_{1\gamma} - q_{1\delta}|) dq_1^N \cdots dq_m^N$$

$$= \frac{1}{N} \sum_{\gamma \neq \delta} \int dq_{1\gamma} dq_{1\delta} f_{\gamma\delta}(|q_{1\gamma} - q_{1\delta}|) \underbrace{\left( \frac{N}{V^N} \int \frac{e^{-\beta U_\gamma}}{Q_{0\gamma}} dq_\gamma^{N-1} \right)}_{\rho^{(1)}(q_{1\gamma})} \cdots \underbrace{\left( \frac{N}{V^N} \int \frac{e^{-\beta U_\delta}}{Q_{0\delta}} dq_\delta^{N-1} \right)}_{\rho^{(1)}(q_{1\delta})} \cdots \underbrace{\left( \frac{1}{V^N} \int \frac{e^{-\beta U_m}}{Q_{0m}} dq_m^N \right)}_{1}$$

and hence:

$$a_{12} = \frac{1}{N} \sum_{\gamma \neq \delta} \int dq_{1\gamma} dq_{1\delta} f_{\gamma\delta}(|q_{1\gamma} - q_{1\delta}|) \rho^{(1)}(q_{1\gamma}) \rho^{(1)}(q_{1\delta}).$$

For fluid systems it becomes:

$$a_{12} = \rho \sum_{\gamma \neq \delta} \int dq' f_{\gamma\delta}(|q'|)$$

Since this term is not extensive and proportional to $N$ it will vanish in the thermodynamic limit. This means that intra-potentials do not have contributions to the excess free energy of the system at the first level of approximation in perturbation theory. In what follows it will be very useful to employ the Fourier Transform of a function and its inverse defined by $\tilde{f}(k) = \int f(r) e^{-i\mathbf{k}\cdot\mathbf{r}} d^3 r$ and $f(r) = \frac{1}{(2\pi)^3} \int \tilde{f}(k) e^{i\mathbf{k}\cdot\mathbf{r}} d^3 k$.

The reduced free energy of a molecular fluid will thus be found up to the first order perturbation theory as:

$$a = \sum_{\mu} a_{0\mu} - \frac{\rho}{2} \sum_{\mu \neq \nu} \tilde{f}_{\mu\nu}(0) \tag{24}$$

Although this equation looks good and may be employed to obtain the thermodynamic of molecular liquids, but as one can simply guess we can proceed further. We will thus calculate the second term contribution in the hope of achieving better thermodynamics.

**4. Calculation of the Second Perturbation Contributions to the Reduced Free Energy of the System**

Calculation of $a_2$, though not an easy task, is also possible. It comprises of three contributions that are calculated in the following subsections.

**4.1. Calculation of $a_{21}$**

The explicit form of $a_{21}$ has been given in Eqs. (16), (17), and (21) which reads:

$$a_{21} = \frac{1}{V^{Nm}} \sum_{\substack{i<j \\ \mu \neq v}} \sum_{\substack{i'<j' \\ \mu' \neq v'}} \int \frac{e^{-\beta \sum_{\alpha=1}^{m} U_\alpha}}{Q_0} f_{\mu v}(|q_{i\mu} - q_{jv}|) f_{\mu' v'}(|q'_{i'\mu'} - q'_{j'v'}|) dq_1^N ... dq_m^N$$

(25)

$$- \frac{1}{V^{Nm}} \frac{1}{V^{Nm}} \sum_{\substack{i<j \\ \mu \neq v}} \sum_{\substack{i'<j' \\ \mu' \neq v'}} \int \frac{e^{-\beta \sum_{\alpha=1}^{m} U_\alpha}}{Q_0} \frac{e^{-\beta \sum_{\alpha=1}^{m} U_{\alpha'}}}{Q_0} f_{\mu v}(|q_{i\mu} - q_{jv}|) f_{\mu' v'}(|q'_{i'\mu'} - q'_{j'v'}|) dq_1^N ... dq_m^N dq_1'^N ... dq_m'^N$$

We can distinguish four cases from molecular indexes:

1. $i = i'$, $j = j'$; 2. $i \neq i'$, $j = j'$; 3. $i = i'$, $j \neq j'$; and 4. $i \neq i'$, $j \neq j'$.

To each of which there are four cases from atomic indexes:

1. $\mu = \mu'$, $v = v'$; 2. $\mu \neq \mu'$, $v = v'$; 3. $\mu = \mu'$, $v \neq v'$; and 4. $\mu \neq \mu'$, $v \neq v'$.

Let us calculate them case by case.

11. In this case we have $i = i'$, $j = j'$ and $\mu = \mu'$, $v = v'$. Eq. (25) will then become:

$$a_{21,11} = \frac{1}{V^{Nm}} \sum_{\substack{i<j \\ \mu \neq v}} \int \frac{e^{-\beta \sum_{\alpha=1}^{m} U_\alpha}}{Q_0} f_{\mu v}(|q_{i\mu} - q_{jv}|) f_{\mu v}(|q_{i\mu} - q_{jv}|) dq_1^N ... dq_m^N$$

$$- \frac{1}{V^{Nm}} \frac{1}{V^{Nm}} \sum_{\substack{i<j \\ \mu \neq v}} \int \frac{e^{-\beta \sum_{\alpha=1}^{m} U_\alpha}}{Q_0} \frac{e^{-\beta \sum_{\alpha=1}^{m} U_{\alpha'}}}{Q_0} f_{\mu v}(|q_{i\mu} - q_{jv}|) f_{\mu v}(|q'_{i\mu} - q'_{jv}|) dq_1^N ... dq_m^N dq_1'^N ... dq_m'^N$$

which is equal to:

$$a_{21,11} = \frac{N(N-1)}{2} \sum_{\mu \neq v} \int f_{\mu v}(|q_{1\mu} - q_{2v}|) f_{\mu v}(|q_{1\mu} - q_{2v}|) dq_{1\mu} dq_{2v} \times \frac{\rho^2}{N^2}$$

$$- \frac{N(N-1)}{2} \sum_{\mu \neq v} \int f_{\mu v}(|q_{1\mu} - q_{2v}|) dq_{1\mu} dq_{2v} \times \frac{\rho^2}{N^2} \int f_{\mu v}(|q'_{1\mu} - q'_{2v}|) dq'_{1\mu} dq'_{2v} \times \frac{\rho^2}{N^2}$$

The second term vanishes in thermodynamic limit and in the first term by taking the variable $q_{2v}$ out of the integral sign we will find:

$$a_{21,11} = \frac{\rho N}{2} \sum_{\mu \neq \nu} \int f_{\mu\nu}^2(|q'_{1\mu}|) dq'_{1\mu}$$

which is proportional to $N$. By using Fourier transform methods it becomes:

$$a_{21,11} = \frac{\rho N}{2} \sum_{\mu \neq \nu} \widetilde{f_{\mu\nu}^2}(0) \quad (26)$$

12. In this case we have $i = i'$, $j = j'$ and $\mu \neq \mu'$, $\nu = \nu'$. Eq. (25) will then become:

$$a_{21,12} = \frac{1}{V^{Nm}} \sum_{\substack{i<j \\ \mu \neq \nu}} \sum_{\mu'} \int \frac{e^{-\beta \sum_{\alpha=1}^{m} U_\alpha}}{Q_0} f_{\mu\nu}(|q_{i\mu} - q_{j\nu}|) f_{\mu'\nu}(|q_{i\mu'} - q_{j\nu}|) dq_1^N ... dq_m^N$$

$$- \frac{1}{V^{Nm}} \frac{1}{V^{Nm}} \sum_{\substack{i<j \\ \mu \neq \nu}} \sum_{\mu'} \int \frac{e^{-\beta \sum_{\alpha=1}^{m} U_\alpha}}{Q_0} \frac{e^{-\beta \sum_{\alpha=1}^{m} U_{\alpha'}}}{Q_0} f_{\mu\nu}(|q_{i\mu} - q_{j\nu}|) f_{\mu'\nu}(|q'_{i\mu'} - q'_{j\nu}|) dq_1^N ... dq_m^N dq_1^{'N} ... dq_m^{'N}$$

This is simply equal to:

$$a_{21,12} = \frac{N(N-1)}{2} \sum_{\mu \neq \nu} \sum_{\mu'} \int f_{\mu\nu}(|q_{1\mu} - q_{2\nu}|) f_{\mu'\nu}(|q_{1\mu'} - q_{2\nu}|) dq_{1\mu} dq_{2\nu} dq_{1\mu'} \times \frac{\rho^3}{N^3}$$

$$- \frac{N(N-1)}{2} \sum_{\mu \neq \nu} \sum_{\mu'} \int f_{\mu\nu}(|q_{1\mu} - q_{2\nu}|) dq_{1\mu} dq_{2\nu} \times \frac{\rho^2}{N^2} \int f_{\mu'\nu}(|q'_{1\mu'} - q'_{2\nu}|) dq'_{1\mu'} dq'_{2\nu} \times \frac{\rho^2}{N^2}$$

or:

$$a_{21,12} = \frac{\rho^3 V}{2N} \sum_{\mu \neq \nu} \sum_{\mu'} \int f_{\mu\nu}(|q_{1\mu}|) f_{\mu'\nu}(|q_{1\mu'}|) dq_{1\mu} dq_{1\mu'}$$

$$- \frac{\rho^4 V^2}{2N^2} \sum_{\mu \neq \nu} \sum_{\mu'} \int f_{\mu\nu}(|q_{1\mu}|) dq_{1\mu} \int f_{\mu'\nu}(|q'_{1\mu'}|) dq'_{1\mu'}$$

that is identically zero.

13. This case with $i = i'$, $j = j'$ and $\mu = \mu'$, $\nu \neq \nu'$ is completely identical to that of case 12 with no contribution.

14. In this case we have $i = i'$, $j = j'$ and $\mu \neq \mu'$, $\nu \neq \nu'$. Eq. (25) will in this case become:

$$a_{21,14} = \frac{1}{V^{Nm}} \sum_{\substack{i<j \\ \mu \neq \nu}} \sum_{\substack{\mu' \neq \nu'}} \int \frac{e^{-\beta \sum_{\alpha=1}^{m} U_\alpha}}{Q_0} f_{\mu\nu}(|q_{i\mu} - q_{j\nu}|) f_{\mu'\nu'}(|q_{i\mu'} - q_{j\nu'}|) dq_1^N ... dq_m^N$$

$$- \frac{1}{V^{Nm}} \frac{1}{V^{Nm}} \sum_{\substack{i<j \\ \mu \neq \nu}} \sum_{\mu' \neq \nu'} \int \frac{e^{-\beta \sum_{\alpha=1}^{m} U_\alpha}}{Q_0} \frac{e^{-\beta \sum_{\alpha=1}^{m} U_{\alpha'}}}{Q_0} f_{\mu\nu}(|q_{i\mu} - q_{j\nu}|) f_{\mu'\nu'}(|q'_{i\mu'} - q'_{j\nu'}|) dq_1^N ... dq_m^N dq_1'^N ... dq_m'^N$$

which is equal to:

$$a_{21,14} = \frac{N(N-1)}{2} \sum_{\mu \neq \nu} \sum_{\mu' \neq \nu'} \int f_{\mu\nu}(|q_{1\mu} - q_{2\nu}|) f_{\mu'\nu'}(|q_{1\mu'} - q_{2\nu'}|) dq_{1\mu} dq_{2\nu} dq_{1\mu'} dq_{2\nu'} \times \frac{\rho^4}{N^4}$$

$$- \frac{N(N-1)}{2} \sum_{\mu \neq \nu} \sum_{\mu' \neq \nu'} \int f_{\mu\nu}(|q_{1\mu} - q_{2\nu}|) dq_{1\mu} dq_{2\nu} \times \frac{\rho^2}{N^2} \int f_{\mu'\nu'}(|q'_{1\mu'} - q'_{2\nu'}|) dq'_{1\mu'} dq'_{2\nu'} \times \frac{\rho^2}{N^2}$$

that is identically zero. So, the first group has only one contribution, Eq. (26).

21. In this case we have $i \neq i'$, $j = j'$ and $\mu = \mu'$, $\nu = \nu'$. Eq. (25) will then become:

$$a_{21,21} = \frac{1}{V^{Nm}} \sum_{\substack{i<j \\ \mu \neq \nu}} \sum_{i'} \int \frac{e^{-\beta \sum_{\alpha=1}^{m} U_\alpha}}{Q_0} f_{\mu\nu}(|q_{i\mu} - q_{j\nu}|) f_{\mu\nu}(|q_{i'\mu} - q_{j\nu}|) dq_1^N ... dq_m^N$$

$$- \frac{1}{V^{Nm}} \frac{1}{V^{Nm}} \sum_{\substack{i<j \\ \mu \neq \nu}} \sum_{i'} \int \frac{e^{-\beta \sum_{\alpha=1}^{m} U_\alpha}}{Q_0} \frac{e^{-\beta \sum_{\alpha=1}^{m} U_{\alpha'}}}{Q_0} f_{\mu\nu}(|q_{i\mu} - q_{j\nu}|) f_{\mu\nu}(|q'_{i'\mu} - q'_{j\nu}|) dq_1^N ... dq_m^N dq_1'^N ... dq_m'^N$$

which equals to:

$$a_{21,21} = \frac{N(N-1)^2}{2} \sum_{\mu \neq \nu} \int f_{\mu\nu}(|q_{1\mu} - q_{2\nu}|) f_{\mu\nu}(|q_{3\mu} - q_{2\nu}|) g_\mu(|q_{1\mu} - q_{3\mu}|) dq_{1\mu} dq_{2\nu} dq_{3\mu} \times \frac{\rho^3}{N^3}$$

$$- \frac{N(N-1)^2}{2} \sum_{\mu \neq \nu} \int f_{\mu\nu}(|q_{1\mu} - q_{2\nu}|) dq_{1\mu} dq_{2\nu} \times \frac{\rho^2}{N^2} \int f_{\mu\nu}(|q'_{3\mu} - q'_{2\nu}|) dq'_{3\mu} dq'_{2\nu} \times \frac{\rho^2}{N^2}$$

or:

$$a_{21,21} = \frac{\rho^3}{2} \sum_{\mu \neq \nu} \int f_{\mu\nu}(|q_{1\mu} - q_{2\nu}|) f_{\mu\nu}(|q_{3\mu} - q_{2\nu}|) g_{\mu}(|q_{1\mu} - q_{3\mu}|) dq_{1\mu} dq_{2\nu} dq_{3\mu}$$

$$- \frac{\rho^4}{2N} \sum_{\mu \neq \nu} \int f_{\mu\nu}(|q_{1\mu} - q_{2\nu}|) dq_{1\mu} dq_{2\nu} \int f_{\mu\nu}(|q'_{3\mu} - q'_{2\nu}|) dq'_{3\mu} dq'_{2\nu}$$

which may be even simpler if we define new variables $q'_{2\nu} = q_{2\nu} - q_{1\mu}$, $q'_{3\mu} = q_{3\mu} - q_{1\mu}$. By use of these new variables it becomes:

$$a_{21,21} = \frac{\rho^3 V}{2} \sum_{\mu \neq \nu} \int f_{\mu\nu}(|q'_{2\nu}|) f_{\mu\nu}(|q'_{3\mu} - q'_{2\nu}|) g_{\mu}(|q'_{3\mu}|) dq'_{2\nu} dq'_{3\mu} - \frac{\rho^4 V^2}{2N} \sum_{\mu \neq \nu} \int f_{\mu\nu}(|q'_{2\nu}|) dq'_{2\nu} \int f_{\mu\nu}(|q'_{3\mu}|) dq'_{3\mu} d$$

By use of indirect correlation function defined by $h(q) = g(q) - 1$, it simplifies to:

$$a_{21,21} = \frac{\rho^2 N}{2} \sum_{\mu \neq \nu} \int f_{\mu\nu}(|q_{2\nu}|) f_{\mu\nu}(|q_{3\mu} - q_{2\nu}|) h_{\mu}(|q_{3\mu}|) dq_{2\nu} dq_{3\mu}$$

which can be transformed into (by employing Fourier Transform methods):

$$a_{21,21} = \frac{\rho^2 N}{2} \sum_{\mu \neq \nu} \int \tilde{f}^2_{\mu\nu}(k) \tilde{h}_{\mu}(k) \frac{d^3 k}{(2\pi)^3} \qquad (27).$$

As seen it is proportional to $N$, hence will survive in the thermodynamic limit.

22. In this case we have $i \neq i'$, $j = j'$ and $\mu \neq \mu'$, $\nu = \nu'$. Eq. (25) will then become:

$$a_{21,22} = \frac{1}{V^{Nm}} \sum_{\substack{i<j \\ \mu,\nu}} \sum_{\substack{i' \\ \mu'}} \int \frac{e^{-\beta \sum_{\alpha=1}^{m} U_\alpha}}{Q_0} f_{\mu\nu}(|q_{i\mu} - q_{j\nu}|) f_{\mu'\nu}(|q_{i'\mu'} - q_{j\nu}|) dq_1^N ... dq_m^N$$

$$- \frac{1}{V^{Nm}} \frac{1}{V^{Nm}} \sum_{\substack{i<j \\ \mu,\nu}} \sum_{\substack{i' \\ \mu'}} \int \frac{e^{-\beta \sum_{\alpha=1}^{m} U_\alpha}}{Q_0} \frac{e^{-\beta \sum_{\alpha=1}^{m} U_{\alpha'}}}{Q_0} f_{\mu\nu}(|q_{i\mu} - q_{j\nu}|) f_{\mu'\nu}(|q'_{i'\mu'} - q'_{j\nu}|) dq_1^N ... dq_m^N dq'^N_1 ... dq'^N_m$$

This is simply equal to:

$$a_{21,22} = \frac{N(N-1)^2}{2} \sum_{\mu \neq \nu} \sum_{\mu'} \int f_{\mu\nu}(|q_{1\mu} - q_{2\nu}|) f_{\mu'\nu}(|q_{3\mu'} - q_{2\nu}|) dq_{1\mu} dq_{2\nu} dq_{3\mu'} \times \frac{\rho^3}{N^3}$$

$$-\frac{N(N-1)^2}{2} \sum_{\mu \neq \nu} \sum_{\mu'} \int f_{\mu\nu}(|q_{1\mu} - q_{2\nu}|) dq_{1\mu} dq_{2\nu} \times \frac{\rho^2}{N^2} \int f_{\mu'\nu}(|q'_{3\mu'} - q'_{2\nu}|) dq'_{3\mu'} dq'_{2\nu} \times \frac{\rho^2}{N^2}$$

or:

$$a_{21,22} = \frac{\rho^3}{2} \sum_{\mu \neq \nu} \sum_{\mu'} \int f_{\mu\nu}(|q_{1\mu} - q_{2\nu}|) f_{\mu'\nu}(|q_{3\mu'} - q_{2\nu}|) dq_{1\mu} dq_{2\nu} dq_{3\mu'}$$

$$-\frac{\rho^4}{2N} \sum_{\mu \neq \nu} \sum_{\mu'} \int f_{\mu\nu}(|q_{1\mu} - q_{2\nu}|) dq_{1\mu} dq_{2\nu} \int f_{\mu'\nu}(|q'_{3\mu'} - q'_{2\nu}|) dq'_{3\mu'} dq'_{2\nu}$$

that is identically zero.

23. For the case with $i \neq i'$, $j = j'$ and $\mu = \mu'$, $\nu \neq \nu'$, Eq. (25) will become:

$$a_{21,23} = \frac{1}{V^{Nm}} \sum_{\substack{i<j \\ \mu \neq \nu}} \sum_{\substack{i' \\ \nu'}} \int \frac{e^{-\beta \sum_{\alpha=1}^{m} U_\alpha}}{Q_0} f_{\mu\nu}(|q_{i\mu} - q_{j\nu}|) f_{\mu\nu'}(|q_{i'\mu} - q_{j\nu'}|) dq_1^N ... dq_m^N$$

$$-\frac{1}{V^{Nm}} \frac{1}{V^{Nm}} \sum_{\substack{i<j \\ \mu \neq \nu}} \sum_{\substack{i' \\ \nu'}} \int \frac{e^{-\beta \sum_{\alpha=1}^{m} U_\alpha}}{Q_0} \frac{e^{-\beta \sum_{\alpha=1}^{m} U_{\alpha'}}}{Q_0} f_{\mu\nu}(|q_{i\mu} - q_{j\nu}|) f_{\mu\nu'}(|q'_{i'\mu} - q'_{j\nu'}|) dq_1^N ... dq_m^N dq_1'^N ... dq_m'^N$$

That is equal to:

$$a_{21,23} = \frac{N(N-1)^2}{2} \sum_{\mu \neq \nu} \sum_{\nu'} \int f_{\mu\nu}(|q_{1\mu} - q_{2\nu}|) f_{\mu\nu'}(|q_{3\mu} - q_{2\nu'}|) g_\mu(|q_{1\mu} - q_{3\mu}|) dq_{1\mu} dq_{2\nu} dq_{3\mu} dq_{2\nu'} \times \frac{\rho^4}{N^4}$$

$$-\frac{N(N-1)^2}{2} \sum_{\mu \neq \nu} \sum_{\nu'} \int f_{\mu\nu}(|q_{1\mu} - q_{2\nu}|) dq_{1\mu} dq_{2\nu} \times \frac{\rho^2}{N^2} \int f_{\mu\nu'}(|q'_{3\mu} - q'_{2\nu'}|) dq'_{3\mu} dq'_{2\nu'} \times \frac{\rho^2}{N^2}$$

or:

$$a_{21,23} = \frac{\rho^4}{2N} \sum_{\mu \neq \nu} \sum_{\nu'} \int f_{\mu\nu}(|q_{1\mu} - q_{2\nu}|) f_{\mu\nu'}(|q_{3\mu} - q_{2\nu'}|) h_\mu(|q_{1\mu} - q_{3\mu}|) dq_{1\mu} dq_{2\nu} dq_{3\mu} dq_{2\nu'}$$

We can express all variables in terms of $q_{1\mu}$ and take it out of the integral sign to get:

$$a_{21,23} = \frac{\rho^3}{2} \sum_{\substack{\mu \neq \nu \\ \nu'}} \int f_{\mu\nu}(|q'_{2\nu}|) f_{\mu\nu'}(|q'_{3\mu} - q'_{2\nu'}|) h_\mu(|q'_{3\mu}|) dq'_{2\nu} dq'_{3\mu} dq'_{2\nu'} = \frac{\rho^3}{2(2\pi)^3} \sum_{\substack{\mu \neq \nu \\ \nu'}} \tilde{f}_{\mu\nu}(0) \tilde{f}_{\mu\nu'}(0) \tilde{h}_\mu(0)$$

Since the expression is not proportional to $N$ it vanishes in the thermodynamic limit.

24. In this case we have $i \neq i'$, $j = j'$ and $\mu \neq \mu'$, $\nu \neq \nu'$. Eq. (25) will then become:

$$a_{21,24} = \frac{1}{V^{Nm}} \sum_{\substack{i < j \\ \mu \neq \nu}} \sum_{\substack{i' \\ \mu' \neq \nu'}} \int \frac{e^{-\beta \sum_{\alpha=1}^{m} U_\alpha}}{Q_0} f_{\mu\nu}(|q_{1\mu} - q_{2\nu}|) f_{\mu'\nu'}(|q'_{i'\mu'} - q'_{j\nu'}|) dq_1^N ... dq_m^N$$

$$- \frac{1}{V^{Nm}} \frac{1}{V^{Nm}} \sum_{\substack{i < j \\ \mu \neq \nu}} \sum_{\substack{i' \\ \mu' \neq \nu'}} \int \frac{e^{-\beta \sum_{\alpha=1}^{m} U_\alpha}}{Q_0} \frac{e^{-\beta \sum_{\alpha=1}^{m} U_{\alpha'}}}{Q_0} f_{\mu\nu}(|q_{1\mu} - q_{2\nu}|) f_{\mu'\nu'}(|q'_{i'\mu'} - q'_{j\nu'}|) dq_1^N ... dq_m^N dq'^{'N}_1 ... dq'^{'N}_m$$

This is equal to:

$$a_{21,24} = \frac{N(N-1)^2}{2} \sum_{\mu \neq \nu} \sum_{\mu' \neq \nu'} \int f_{\mu\nu}(|q_{1\mu} - q_{2\nu}|) f_{\mu'\nu'}(|q_{3\mu'} - q_{2\nu'}|) dq_{1\mu} dq_{2\nu} dq_{3\mu'} dq_{2\nu'} \times \frac{\rho^4}{N^4}$$

$$- \frac{N(N-1)^2}{2} \sum_{\mu \neq \nu} \sum_{\mu' \neq \nu'} \int f_{\mu\nu}(|q_{1\mu} - q_{2\nu}|) dq_{1\mu} dq_{2\nu} \times \frac{\rho^2}{N^2} \int f_{\mu'\nu'}(|q'_{3\mu'} - q'_{2\nu'}|) dq'_{3\mu'} dq'_{2\nu'} \times \frac{\rho^2}{N^2}$$

that vanishes identically. So, the second group has only one contribution, Eq. (27).

Furthermore, the third group is essentially identical to the second group by interchanging the roles of $i$ and $j$ and thus gives out a contribution essentially the same as Eq. (27), that is:

$$a_{21,31} = \frac{\rho^2 N}{2} \sum_{\mu \neq \nu} \int \tilde{f}_{\mu\nu}^2(k) \tilde{h}_\nu(k) \frac{d^3k}{(2\pi)^3} \tag{28}.$$

Now we start calculating the fourth group cases.

41. In this case we have $i \neq i'$, $j \neq j'$ and $\mu = \mu'$, $\nu = \nu'$. Eq. (25) will then become:

$$a_{21,41} = \frac{1}{V^{Nm}} \sum_{\substack{i<j \\ \mu \neq \nu}} \sum_{i<j} \int \frac{e^{-\beta \sum_{\alpha=1}^{m} U_\alpha}}{Q_0} f_{\mu\nu}(|q_{i\mu} - q_{j\nu}|) f_{\mu\nu}(|q_{i'\mu} - q_{j'\nu}|) dq_1^N ... dq_m^N$$

$$- \frac{1}{V^{Nm}} \frac{1}{V^{Nm}} \sum_{\substack{i<j \\ \mu \neq \nu}} \sum_{i<j} \int \frac{e^{-\beta \sum_{\alpha=1}^{m} U_\alpha}}{Q_0} \frac{e^{-\beta \sum_{\alpha=1}^{m} U_{\alpha'}}}{Q_0} f_{\mu\nu}(|q_{i\mu} - q_{j\nu}|) f_{\mu\nu}(|q'_{i'\mu} - q'_{j'\nu}|) dq_1^N ... dq_m^N dq_1^{'N} ... dq_m^{'N}$$

which equals to:

$$a_{21,41} = \frac{N^2(N-1)^2}{4} \sum_{\mu \neq \nu} \int f_{\mu\nu}(|q_{1\mu} - q_{2\nu}|) f_{\mu\nu}(|q_{3\mu} - q_{4\nu}|) g_\mu(|q_{1\mu} - q_{3\mu}|) g_\nu(|q_{2\nu} - q_{4\nu}|) dq_{1\mu} dq_{2\nu} dq_{3\mu} dq_{4\nu} \times \frac{\rho^4}{N^4}$$

$$- \frac{N^2(N-1)^2}{4} \sum_{\mu \neq \nu} \int f_{\mu\nu}(|q_{1\mu} - q_{2\nu}|) dq_{1\mu} dq_{2\nu} \times \frac{\rho^2}{N^2} \int f_{\mu\nu}(|q'_{3\mu} - q'_{4\nu}|) dq'_{3\mu} dq'_{4\nu} \times \frac{\rho^2}{N^2}$$

This will take a simpler form if we invoke indirect correlation functions:

$$a_{21,41} = \frac{\rho^4}{4} \sum_{\mu \neq \nu} \int f_{\mu\nu}(|q_{1\mu} - q_{2\nu}|) f_{\mu\nu}(|q_{3\mu} - q_{4\nu}|) \left\{ [h_\mu(|q_{1\mu} - q_{3\mu}|) + 1][h_\nu(|q_{2\nu} - q_{4\nu}|) + 1] - 1 \right\} dq_{1\mu} dq_{2\nu} dq_{3\mu} dq_{4\nu}$$

This can be split into three simpler terms:

$$a_{21,41} = \frac{\rho^4}{4} \sum_{\mu \neq \nu} \int f_{\mu\nu}(|q_{1\mu} - q_{2\nu}|) f_{\mu\nu}(|q_{3\mu} - q_{4\nu}|) h_\mu(|q_{1\mu} - q_{3\mu}|) dq_{1\mu} dq_{2\nu} dq_{3\mu} dq_{4\nu}$$

$$+ \frac{\rho^4}{4} \sum_{\mu \neq \nu} \int f_{\mu\nu}(|q_{1\mu} - q_{2\nu}|) f_{\mu\nu}(|q_{3\mu} - q_{4\nu}|) h_\nu(|q_{2\nu} - q_{4\nu}|) dq_{1\mu} dq_{2\nu} dq_{3\mu} dq_{4\nu}$$

$$+ \frac{\rho^4}{4} \sum_{\mu \neq \nu} \int f_{\mu\nu}(|q_{1\mu} - q_{2\nu}|) f_{\mu\nu}(|q_{3\mu} - q_{4\nu}|) h_\mu(|q_{1\mu} - q_{3\mu}|) h_\nu(|q_{2\nu} - q_{4\nu}|) dq_{1\mu} dq_{2\nu} dq_{3\mu} dq_{4\nu}$$

Taking one suitable variable out of each integral yields:

$$a_{21,41} = \frac{\rho^4 V}{4} \sum_{\mu \neq \nu} \int f_{\mu\nu}(|q'_{2\nu}|) dq'_{2\nu} \int f_{\mu\nu}(|q'_{3\mu} - q'_{4\nu}|) h_\mu(|q'_{3\mu}|) dq'_{3\mu} dq'_{4\nu}$$

$$+ \frac{\rho^4 V}{4} \sum_{\mu \neq \nu} \int f_{\mu\nu}(|q'_{1\mu}|) dq'_{1\mu} \int f_{\mu\nu}(|q'_{4\nu} - q'_{3\mu}|) h_\nu(|q'_{4\nu}|) dq'_{3\mu} dq'_{4\nu}$$

$$+ \frac{\rho^4 V}{4} \sum_{\mu \neq \nu} \int f_{\mu\nu}(|q'_{2\nu}|) h_\nu(|q'_{2\nu} - q'_{4\nu}|) f_{\mu\nu}(|q'_{3\mu} - q'_{4\nu}|) h_\mu(|q'_{3\mu}|) dq'_{2\nu} dq'_{3\mu} dq'_{4\nu}$$

As before by use of Fourier Transform methods one finds:

$$a_{21,41} = \frac{\rho^3 N}{4} \sum_{\mu \neq \nu} \left\{ \tilde{f}_{\mu\nu}^2(0)\tilde{h}_\mu(0) + \tilde{f}_{\mu\nu}^2(0)\tilde{h}_\nu(0) + \int \tilde{f}_{\mu\nu}^2(k)\tilde{h}_\mu(k)\tilde{h}_\nu(k)\frac{d^3k}{(2\pi)^3} \right\} \tag{29}$$

42. In this case we have $i \neq i'$, $j \neq j'$ and $\mu \neq \mu'$, $\nu = \nu'$. Eq. (25) will then become:

$$a_{21,42} = \frac{1}{V^{Nm}} \sum_{\substack{i<j \\ \mu \neq \nu}} \sum_{\substack{i'<j' \\ \mu'}} \int \frac{e^{-\beta \sum_{\alpha=1}^{m} U_\alpha}}{Q_0} f_{\mu\nu}(|q_{i\mu} - q_{j\nu}|) f_{\mu'\nu}(|q_{i'\mu'} - q_{j'\nu}|) dq_1^N ... dq_m^N$$

$$-\frac{1}{V^{Nm}} \frac{1}{V^{Nm}} \sum_{\substack{i<j \\ \mu \neq \nu}} \sum_{\substack{i'<j' \\ \mu'}} \int \frac{e^{-\beta \sum_{\alpha=1}^{m} U_\alpha}}{Q_0} \frac{e^{-\beta \sum_{\alpha=1}^{m} U_{\alpha'}}}{Q_0} f_{\mu\nu}(|q_{i\mu} - q_{j\nu}|) f_{\mu'\nu}(|q'_{i'\mu'} - q'_{j'\nu}|) dq_1^N ... dq_m^N dq'^N_1 ... dq'^N_m$$

This is simply equal to:

$$a_{21,42} = \frac{N^2(N-1)^2}{4} \sum_{\substack{\mu \neq \nu \\ \mu'}} \int f_{\mu\nu}(|q_{1\mu} - q_{2\nu}|) f_{\mu'\nu}(|q_{3\mu'} - q_{4\nu}|) g_\nu(|q_{2\nu} - q_{4\nu}|) dq_{1\mu} dq_{2\nu} dq_{3\mu'} dq_{4\nu} \times \frac{\rho^4}{N^4}$$

$$-\frac{N^2(N-1)^2}{4} \sum_{\substack{\mu \neq \nu \\ \mu'}} \int f_{\mu\nu}(|q_{1\mu} - q_{2\nu}|) dq_{1\mu} dq_{2\nu} \times \frac{\rho^2}{N^2} \int f_{\mu'\nu}(|q'_{3\mu'} - q'_{4\nu}|) dq'_{3\mu'} dq'_{4\nu} \times \frac{\rho^2}{N^2}$$

this is equal to:

$$a_{21,42} = \frac{\rho^4}{4} \sum_{\substack{\mu \neq \nu \\ \mu'}} \int f_{\mu\nu}(|q_{1\mu} - q_{2\nu}|) f_{\mu'\nu}(|q_{3\mu'} - q_{4\nu}|) h_\nu(|q_{2\nu} - q_{4\nu}|) dq_{1\mu} dq_{2\nu} dq_{3\mu'} dq_{4\nu}$$

or equivalently:

$$a_{21,42} = \frac{\rho^3 N}{4} \sum_{\substack{\mu \neq \nu \\ \mu'}} \tilde{f}_{\mu\nu}(0)\tilde{f}_{\mu'\nu}(0)\tilde{h}_\nu(0) \tag{30}.$$

43. In this case we have $i \neq i'$, $j \neq j'$ and $\mu = \mu'$, $\nu \neq \nu'$. Eq. (25) will then become:

$$a_{21,43} = \frac{1}{V^{Nm}} \sum_{\substack{i<j \\ \mu \neq v}} \sum_{\substack{i'<j' \\ v'}} \int \frac{e^{-\beta \sum_{\alpha=1}^{m} U_\alpha}}{Q_0} f_{\mu v}(|q_{i\mu} - q_{jv}|) f_{\mu v'}(|q_{i'\mu} - q_{j'v'}|) dq_1^N ... dq_m^N$$

$$- \frac{1}{V^{Nm}} \frac{1}{V^{Nm}} \sum_{\substack{i<j \\ \mu \neq v}} \sum_{\substack{i'<j' \\ v'}} \int \frac{e^{-\beta \sum_{\alpha=1}^{m} U_\alpha}}{Q_0} \frac{e^{-\beta \sum_{\alpha=1}^{m} U_{\alpha'}}}{Q_0} f_{\mu v}(|q_{i\mu} - q_{jv}|) f_{\mu v'}(|q'_{i'\mu} - q'_{j'v'}|) dq_1^N ... dq_m^N dq'_1^{'N} ... dq'_m^{'N}$$

This is simply equal to:

$$a_{21,43} = \frac{N^2(N-1)^2}{4} \sum_{\substack{\mu \neq v \\ v'}} \int f_{\mu v}(|q_{1\mu} - q_{2v}|) f_{\mu v'}(|q_{3\mu} - q_{4v'}|) g_\mu(|q_{1\mu} - q_{3\mu}|) dq_{1\mu} dq_{2v} dq_{3\mu} dq_{4v'} \times \frac{\rho^4}{N^4}$$

$$- \frac{N^2(N-1)^2}{4} \sum_{\substack{\mu \neq v \\ v'}} \int f_{\mu v}(|q_{1\mu} - q_{2v}|) dq_{1\mu} dq_{2v} \times \frac{\rho^2}{N^2} \int f_{\mu v'}(|q'_{3\mu} - q'_{4v'}|) dq'_{3\mu} dq'_{4v'} \times \frac{\rho^2}{N^2}$$

which is equal to:

$$a_{21,42} = \frac{\rho^4}{4} \sum_{\substack{\mu \neq v \\ v'}} \int f_{\mu v}(|q_{1\mu} - q_{2v}|) f_{\mu v'}(|q_{3\mu} - q_{4v'}|) h_\mu(|q_{1\mu} - q_{3\mu}|) dq_{1\mu} dq_{2v} dq_{3\mu} dq_{4v'}$$

that is equal to:

$$a_{21,43} = \frac{\rho^3 N}{4} \sum_{\substack{\mu \neq v \\ v'}} \tilde{f}_{\mu v}(0) \tilde{f}_{\mu v'}(0) \tilde{h}_\mu(0) \tag{31}.$$

44. In this case we have $i \neq i'$, $j \neq j'$ and $\mu \neq \mu'$, $v \neq v'$. Eq. (25) will then become:

$$a_{21,44} = \frac{1}{V^{Nm}} \sum_{\substack{i<j \\ \mu \neq v}} \sum_{\substack{i'<j' \\ \mu' \neq v'}} \int \frac{e^{-\beta \sum_{\alpha=1}^{m} U_\alpha}}{Q_0} f_{\mu v}(|q_{i\mu} - q_{jv}|) f_{\mu' v'}(|q_{i'\mu'} - q_{j'v'}|) dq_1^N ... dq_m^N$$

$$- \frac{1}{V^{Nm}} \frac{1}{V^{Nm}} \sum_{\substack{i<j \\ \mu \neq v}} \sum_{\substack{i'<j' \\ \mu' \neq v'}} \int \frac{e^{-\beta \sum_{\alpha=1}^{m} U_\alpha}}{Q_0} \frac{e^{-\beta \sum_{\alpha=1}^{m} U_{\alpha'}}}{Q_0} f_{\mu v}(|q_{i\mu} - q_{jv}|) f_{\mu' v'}(|q'_{i'\mu'} - q'_{j'v'}|) dq_1^N ... dq_m^N dq'_1^{'N} ... dq'_m^{'N}$$

This is equal to:

$$a_{21,44} = \frac{N^2(N-1)^2}{4} \sum_{\substack{\mu \neq \nu \\ \mu' \neq \nu'}} \int f_{\mu\nu}(|q_{1\mu} - q_{2\nu}|) f_{\mu'\nu'}(|q_{3\mu'} - q_{4\nu'}|) dq_{1\mu} dq_{2\nu} dq_{3\mu'} dq_{4\nu'} \times \frac{\rho^4}{N^4}$$

$$-\frac{N^2(N-1)^2}{4} \sum_{\substack{\mu \neq \nu \\ \mu' \neq \nu'}} \int f_{\mu\nu}(|q_{1\mu} - q_{2\nu}|) dq_{1\mu} dq_{2\nu} \times \frac{\rho^2}{N^2} \int f_{\mu'\nu'}(|q'_{3\mu'} - q'_{4\nu'}|) dq'_{3\mu'} dq'_{4\nu'} \times \frac{\rho^2}{N^2}$$

that is identically zero.

**4.2. Calculation of $a_{22}$**

The explicit form of $a_{22}$ has been given in Eqs. (16), (18), and (21) which reads:

$$a_{22} = \frac{1}{V^{Nm}} \sum_{\substack{i<j \ k \\ \mu \neq \nu \ \gamma \neq \delta}} \int \frac{e^{-\beta \sum_{\alpha=1}^{m} U_\alpha}}{Q_0} f_{\mu\nu}(|q_{i\mu} - q_{j\nu}|) f_{\gamma\delta}(|q_{k\gamma} - q_{k\delta}|) dq_1^N ... dq_m^N$$

$$-\frac{1}{V^{Nm}} \frac{1}{V^{Nm}} \sum_{\substack{i<j \ k \\ \mu \neq \nu \ \gamma \neq \delta}} \int \frac{e^{-\beta \sum_{\alpha=1}^{m} U_\alpha}}{Q_0} \frac{e^{-\beta \sum_{\alpha=1}^{m} U_{\alpha'}}}{Q_0} f_{\mu\nu}(|q_{i\mu} - q_{j\nu}|) f_{\gamma\delta}(|q'_{k\gamma} - q'_{k\delta}|) dq_1^N ... dq_m^N dq_1'^N ... dq_m'^N$$

(32).

We can distinguish three cases from molecular indexes:

1. $i = k \neq j$; 2. $i \neq j = k$; and 3. $i \neq j \neq k$.

To each of which there are four cases from atomic indexes:

1. $\mu = \gamma$ and $\nu = \delta$; 2. $\mu \neq \gamma$ and $\nu = \delta$; 3. $\mu = \gamma$ and $\nu \neq \delta$; and 4. $\mu \neq \gamma$ and $\nu \neq \delta$.

Let us calculate all these case by case.

11. In this case we have $i = k \neq j$ and $\mu = \gamma$ and $\nu = \delta$. Eq. (32) will then become:

$$a_{22,11} = \frac{1}{V^{Nm}} \sum_{\substack{i<j \\ \mu \neq \nu}} \int \frac{e^{-\beta \sum_{\alpha=1}^{m} U_\alpha}}{Q_0} f_{\mu\nu}(|q_{i\mu} - q_{j\nu}|) f_{\mu\nu}(|q_{i\mu} - q_{i\nu}|) dq_1^N ... dq_m^N$$

$$- \frac{1}{V^{Nm}} \frac{1}{V^{Nm}} \sum_{\substack{i<j \\ \mu \neq \nu}} \int \frac{e^{-\beta \sum_{\alpha=1}^{m} U_\alpha}}{Q_0} \frac{e^{-\beta \sum_{\alpha=1}^{m} U_{\alpha'}}}{Q_0} f_{\mu\nu}(|q_{i\mu} - q_{j\nu}|) f_{\mu\nu}(|q'_{i\mu} - q'_{i\nu}|) dq_1^N ... dq_m^N dq_1^{'N} ... dq_m^{'N}$$

This is simply equal to:

$$a_{22,11} = \frac{N(N-1)}{2} \sum_{\mu \neq \nu} \int f_{\mu\nu}(|q_{1\mu} - q_{2\nu}|) f_{\mu\nu}(|q_{1\mu} - q_{1\nu}|) g_\nu(|q_{1\nu} - q_{2\nu}|) dq_{1\mu} dq_{2\nu} dq_{1\nu} \times \frac{\rho^3}{N^3}$$

$$- \frac{N(N-1)}{2} \sum_{\mu \neq \nu} \int f_{\mu\nu}(|q_{1\mu} - q_{2\nu}|) dq_{1\mu} dq_{2\nu} \times \frac{\rho^2}{N^2} \int f_{\mu\nu}(|q'_{1\mu} - q'_{1\nu}|) dq'_{1\mu} dq'_{1\nu} \times \frac{\rho^2}{N^2}$$

which is equal to:

$$a_{22,11} = \frac{\rho^3 V}{2N} \sum_{\mu \neq \nu} \int f_{\mu\nu}(|q'_{2\nu}|) f_{\mu\nu}(|q'_{1\nu}|) g_\nu(|q'_{1\nu} - q'_{2\nu}|) dq'_{2\nu} dq'_{1\nu} - \frac{\rho^4 V^2}{2N^2} \sum_{\mu \neq \nu} \int f_{\mu\nu}(|q'_{2\nu}|) f_{\mu\nu}(|q'_{1\nu}|) dq'_{2\nu} dq'_{1\nu}$$

It can be made simpler:

$$a_{22,11} = \frac{\rho^2}{2} \sum_{\mu \neq \nu} \int f_{\mu\nu}(|q'_{2\nu}|) f_{\mu\nu}(|q'_{1\nu}|) h_\nu(|q'_{1\nu} - q'_{2\nu}|) dq'_{1\nu} dq'_{2\nu}$$

Let us define for internal pair potential between two bound atoms in one molecule alone:

$$b_{\mu\nu}(k) = \int \left( e^{-\beta u_{\mu\nu}(|q|)} - 1 \right) e^{-k \cdot q} dq = \int f_{\mu\nu}(|q|) e^{-k \cdot q} dq \qquad (33).$$

We deliberately denote it with $b_{\mu\nu}(k)$ in order to distinguish it from $\tilde{f}_{\mu\nu}(k)$ denoting the same integral for two atoms in two different molecules. Use of this definition yields

$$a_{22,11} = \frac{\rho^2}{2} \sum_{\mu \neq \nu} \int b_{\mu\nu}(k) \tilde{f}_{\mu\nu}(k) \tilde{h}_\nu(k) \frac{d^3 k}{(2\pi)^3}.$$ But this vanishes in the thermodynamic limit.

12. In this case we have $i = k \neq j$ and $\mu \neq \gamma$ and $\nu = \delta$. Eq. (32) will then become:

$$a_{22,12} = \frac{1}{V^{Nm}} \sum_{\substack{i<j \\ \mu \neq \nu, \gamma}} \int \frac{e^{-\beta \sum_{\alpha=1}^{m} U_\alpha}}{Q_0} f_{\mu\nu}(|q_{i\mu} - q_{j\nu}|) f_{\gamma\nu}(|q_{i\gamma} - q_{i\nu}|) dq_1^N ... dq_m^N$$

$$- \frac{1}{V^{Nm}} \frac{1}{V^{Nm}} \sum_{\substack{i<j \\ \mu \neq \nu, \gamma}} \int \frac{e^{-\beta \sum_{\alpha=1}^{m} U_\alpha}}{Q_0} \frac{e^{-\beta \sum_{\alpha=1}^{m} U'_\alpha}}{Q_0} f_{\mu\nu}(|q_{i\mu} - q_{j\nu}|) f_{\gamma\nu}(|q'_{i\gamma} - q'_{i\nu}|) dq_1^N ... dq_m^N dq_1'^N ... dq_m'^N$$

This is simply equal to:

$$a_{22,12} = \frac{N(N-1)}{2} \sum_{\mu \neq \nu, \gamma} \int f_{\mu\nu}(|q_{1\mu} - q_{2\nu}|) f_{\gamma\nu}(|q_{1\gamma} - q_{1\nu}|) g_\nu(|q_{1\nu} - q_{2\nu}|) dq_{1\mu} dq_{2\nu} dq_{1\gamma} dq_{1\nu} \times \frac{\rho^4}{N^4}$$

$$- \frac{N(N-1)}{2} \sum_{\mu \neq \nu, \gamma} \int f_{\mu\nu}(|q_{1\mu} - q_{2\nu}|) dq_{1\mu} dq_{2\nu} \times \frac{\rho^2}{N^2} \int f_{\gamma\nu}(|q'_{1\gamma} - q'_{1\nu}|) dq'_{1\gamma} dq'_{1\nu} \times \frac{\rho^2}{N^2}$$

which is equal to:

$$a_{22,12} = \frac{\rho^4}{2N^2} \sum_{\mu \neq \nu, \gamma} \int f_{\mu\nu}(|q_{1\mu} - q_{2\nu}|) f_{\gamma\nu}(|q_{1\gamma} - q_{1\nu}|) h_\nu(|q_{1\nu} - q_{2\nu}|) dq_{1\mu} dq_{2\nu} dq_{1\gamma} dq_{1\nu}$$

This equation may become simpler:

$$a_{22,12} = \frac{\rho^4 V}{2N^2} \sum_{\mu \neq \nu, \gamma} \int f_{\mu\nu}(|q'_{2\nu}|) f_{\gamma\nu}(|q'_{1\gamma} - q'_{1\nu}|) h_\nu(|q'_{1\nu} - q'_{2\nu}|) dq'_{2\nu} dq'_{1\gamma} dq'_{1\nu}$$

This is equal to $a_{22,12} = \frac{\rho^3}{2N} \sum_{\mu \neq \nu, \gamma} b_{\gamma\nu}(0) \tilde{f}_{\mu\nu}(0) \tilde{h}_\nu(0)$ that vanishes in the thermodynamic limit.

13. In this case we have $i = k \neq j$ and $\mu = \gamma$ and $\nu \neq \delta$. Eq. (32) will then become:

$$a_{22,13} = \frac{1}{V^{Nm}} \sum_{\substack{i<j \\ \mu \neq \nu, \delta}} \int \frac{e^{-\beta \sum_{\alpha=1}^{m} U_\alpha}}{Q_0} f_{\mu\nu}(|q_{i\mu} - q_{j\nu}|) f_{\mu\delta}(|q_{i\mu} - q_{i\delta}|) dq_1^N ... dq_m^N$$

$$- \frac{1}{V^{Nm}} \frac{1}{V^{Nm}} \sum_{\substack{i<j \\ \mu \neq \nu, \delta}} \int \frac{e^{-\beta \sum_{\alpha=1}^{m} U_\alpha}}{Q_0} \frac{e^{-\beta \sum_{\alpha=1}^{m} U'_\alpha}}{Q_0} f_{\mu\nu}(|q_{i\mu} - q_{j\nu}|) f_{\mu\delta}(|q'_{i\mu} - q'_{i\delta}|) dq_1^N ... dq_m^N dq_1'^N ... dq_m'^N$$

That is equal to:

$$a_{22,13} = \frac{N(N-1)}{2} \sum_{\mu \neq \nu, \delta} \int f_{\mu\nu}(|q_{1\mu} - q_{2\nu}|) f_{\mu\delta}(|q_{1\mu}' - q_{1\delta}'|) dq_{1\mu} dq_{2\nu} dq_{1\delta}' \times \frac{\rho^3}{N^3}$$

$$-\frac{N(N-1)}{2} \sum_{\mu \neq \nu, \delta} \int f_{\mu\nu}(|q_{1\mu} - q_{2\nu}|) dq_{1\mu} dq_{2\nu} \times \frac{\rho^2}{N^2} \int f_{\mu\delta}(|q_{1\mu}' - q_{1\delta}'|) dq_{1\mu}' dq_{1\delta}' \times \frac{\rho^2}{N^2}$$

which is identically zero.

14. In this case we have $i = k \neq j$ and $\mu \neq \gamma$ and $\nu \neq \delta$. Eq. (32) will then become:

$$a_{22,14} = \frac{1}{V^{Nm}} \sum_{\substack{i<j \\ \mu \neq \nu \\ \gamma \neq \delta}} \int \frac{e^{-\beta \sum_{\alpha=1}^{m} U_\alpha}}{Q_0} f_{\mu\nu}(|q_{i\mu} - q_{j\nu}|) f_{\gamma\delta}(|q_{i\gamma} - q_{i\delta}|) dq_1^N ... dq_m^N$$

$$-\frac{1}{V^{Nm}} \frac{1}{V^{Nm}} \sum_{\substack{i<j \\ \mu \neq \nu \\ \gamma \neq \delta}} \int \frac{e^{-\beta \sum_{\alpha=1}^{m} U_\alpha}}{Q_0} \frac{e^{-\beta \sum_{\alpha=1}^{m} U_{\alpha}'}}{Q_0} f_{\mu\nu}(|q_{i\mu} - q_{j\nu}|) f_{\gamma\delta}(|q_{i\gamma}' - q_{i\delta}'|) dq_1^N ... dq_m^N dq_1^{'N} ... dq_m^{'N}$$

This is simply equal to:

$$a_{22,14} = \frac{N(N-1)}{2} \sum_{\substack{\mu \neq \nu \\ \gamma \neq \delta}} \int f_{\mu\nu}(|q_{1\mu} - q_{2\nu}|) f_{\gamma\delta}(|q_{1\gamma} - q_{1\delta}|) dq_{1\mu} dq_{2\nu} dq_{1\gamma} dq_{1\delta} \times \frac{\rho^4}{N^4}$$

$$-\frac{N(N-1)}{2} \sum_{\substack{\mu \neq \nu \\ \gamma \neq \delta}} \int f_{\mu\nu}(|q_{1\mu} - q_{2\nu}|) dq_{1\mu} dq_{2\nu} \times \frac{\rho^2}{N^2} \int f_{\gamma\delta}(|q_{1\gamma}' - q_{1\delta}'|) dq_{1\gamma}' dq_{1\delta}' \times \frac{\rho^2}{N^2}$$

that is identically zero.

So, the first group of this type has no contribution, and this is also true for the second group that is essentially identical to it. The third group is now calculated.

31. In this case we have $i \neq j \neq k$ and $\mu = \gamma$ and $\nu = \delta$. Eq. (32) will then become:

$$a_{22,31} = \frac{1}{V^{Nm}} \sum_{\substack{i<j \\ \mu \neq \nu}} \sum_{k} \int \frac{e^{-\beta \sum_{\alpha=1}^{m} U_\alpha}}{Q_0} f_{\mu\nu}(|q_{i\mu} - q_{j\nu}|) f_{\mu\nu}(|q_{k\mu} - q_{k\nu}|) dq_1^N ... dq_m^N$$

$$- \frac{1}{V^{Nm}} \frac{1}{V^{Nm}} \sum_{\substack{i<j \\ \mu \neq \nu}} \sum_{k} \int \frac{e^{-\beta \sum_{\alpha=1}^{m} U_\alpha}}{Q_0} \frac{e^{-\beta \sum_{\alpha=1}^{m} U_{\alpha'}}}{Q_0} f_{\mu\nu}(|q_{i\mu} - q_{j\nu}|) f_{\mu\nu}(|q'_{k\mu} - q'_{k\nu}|) dq_1^N ... dq_m^N dq_1^{'N} ... dq_m^{'N}$$

This is simply equal to:

$$a_{22,31} = \frac{N(N-1)(N-2)}{2} \sum_{\mu \neq \nu} \int f_{\mu\nu}(|q_{1\mu} - q_{2\nu}|) f_{\mu\nu}(|q_{3\mu} - q_{3\nu}|) g_\mu(|q_{1\mu} - q_{3\mu}|) g_\nu(|q_{2\nu} - q_{3\nu}|) dq_{1\mu} dq_{2\nu} dq_{3\mu} dq_{3\nu} \times \frac{\rho^4}{N^4}$$

$$- \frac{N(N-1)(N-2)}{2} \sum_{\mu \neq \nu} \int f_{\mu\nu}(|q_{1\mu} - q_{2\nu}|) dq_{1\mu} dq_{2\nu} \times \frac{\rho^2}{N^2} \int f_{\mu\nu}(|q'_{3\mu} - q'_{3\nu}|) dq'_{3\mu} dq'_{3\nu} \times \frac{\rho^2}{N^2}$$

By use of indirect correlation functions it will become:

$$a_{22,31} = \frac{\rho^4}{2N} \sum_{\mu \neq \nu} \int f_{\mu\nu}(|q_{1\mu} - q_{2\nu}|) f_{\mu\nu}(|q_{3\mu} - q_{3\nu}|) \{[h_\mu(|q_{1\mu} - q_{3\mu}|) + 1][h_\nu(|q_{2\nu} - q_{3\nu}|) + 1] - 1\} dq_{1\mu} dq_{2\nu} dq_{3\mu} dq_{3\nu}$$

which can be split into three simpler terms:

$$a_{22,31} = \frac{\rho^4}{2N} \sum_{\mu \neq \nu} \int f_{\mu\nu}(|q_{1\mu} - q_{2\nu}|) f_{\mu\nu}(|q_{3\mu} - q_{3\nu}|) h_\mu(|q_{1\mu} - q_{3\mu}|) dq_{1\mu} dq_{2\nu} dq_{3\mu} dq_{3\nu}$$

$$+ \frac{\rho^4}{2N} \sum_{\mu \neq \nu} \int f_{\mu\nu}(|q_{1\mu} - q_{2\nu}|) f_{\mu\nu}(|q_{3\mu} - q_{3\nu}|) h_\nu(|q_{2\nu} - q_{3\nu}|) dq_{1\mu} dq_{2\nu} dq_{3\mu} dq_{3\nu}$$

$$+ \frac{\rho^4}{2N} \sum_{\mu \neq \nu} \int f_{\mu\nu}(|q_{1\mu} - q_{2\nu}|) f_{\mu\nu}(|q_{3\mu} - q_{3\nu}|) h_\mu(|q_{1\mu} - q_{3\mu}|) h_\nu(|q_{2\nu} - q_{3\nu}|) dq_{1\mu} dq_{2\nu} dq_{3\mu} dq_{3\nu}$$

or

$$a_{22,31} = \frac{\rho^3}{2} \sum_{\mu \neq \nu} \int f_{\mu\nu}(|q'_{3\nu}|) dq'_{3\nu} \int f_{\mu\nu}(|q'_{1\mu} - q'_{2\nu}|) h_\mu(|q'_{1\mu}|) dq'_{1\mu} dq'_{2\nu}$$

$$+ \frac{\rho^3}{2} \sum_{\mu \neq \nu} \int f_{\mu\nu}(|q'_{3\mu}|) dq'_{3\mu} \int f_{\mu\nu}(|q'_{1\mu} - q'_{2\nu}|) h_\nu(|q'_{2\nu}|) dq'_{1\mu} dq'_{2\nu}$$

$$+ \frac{\rho^3}{2} \sum_{\mu \neq \nu} \int f_{\mu\nu}(|q'_{3\nu}|) h_\nu(|q'_{2\nu} - q'_{3\nu}|) f_{\mu\nu}(|q'_{1\mu} - q'_{2\nu}|) h_\mu(|q'_{1\mu}|) dq'_{1\mu} dq'_{3\nu} dq'_{2\nu}$$

One finally gets:

$$a_{22,31} = \frac{\rho^3}{2} \sum_{\mu \neq \nu} \left( b_{\mu\nu}(0) \tilde{f}_{\mu\nu}(0) \tilde{h}_\mu(0) + b_{\mu\nu}(0) \tilde{f}_{\mu\nu}(0) \tilde{h}_\nu(0) + \int b_{\mu\nu}(k) \tilde{f}_{\mu\nu}(k) \tilde{h}_\mu(k) \tilde{h}_\nu(k) \frac{d^3 k}{(2\pi)^3} \right)$$

But it is not proportional to $N$, and hence vanishes in the thermodynamic limit.

32. In this case we have $i \neq j \neq k$ and $\mu \neq \gamma$ and $\nu = \delta$. Eq. (32) will then become:

$$a_{22,32} = \frac{1}{V^{Nm}} \sum_{\substack{i<j \\ \mu \neq \nu}} \sum_{k,\gamma} \int \frac{e^{-\beta \sum_{\alpha=1}^{m} U_\alpha}}{Q_0} f_{\mu\nu}(|q_{i\mu} - q_{j\nu}|) f_{\gamma\nu}(|q_{k\gamma} - q_{k\nu}|) dq_1^N \ldots dq_m^N$$

$$- \frac{1}{V^{Nm}} \frac{1}{V^{Nm}} \sum_{\substack{i<j \\ \mu \neq \nu}} \sum_{k,\gamma} \int \frac{e^{-\beta \sum_{\alpha=1}^{m} U_\alpha}}{Q_0} \frac{e^{-\beta \sum_{\alpha=1}^{m} U_{\alpha'}}}{Q_0} f_{\mu\nu}(|q_{i\mu} - q_{j\nu}|) f_{\gamma\nu}(|q'_{k\gamma} - q'_{k\nu}|) dq_1^N \ldots dq_m^N dq_1'^N \ldots dq_m'^N$$

This is simply equal to:

$$a_{22,32} = \frac{N(N-1)(N-2)}{2} \sum_{\mu \neq \nu, \gamma} \int f_{\mu\nu}(|q_{1\mu} - q_{2\nu}|) f_{\gamma\nu}(|q_{3\gamma} - q_{3\nu}|) g_\nu(|q_{2\beta\nu} - q_{3\nu}|) dq_{1\mu} dq_{2\nu} dq_{3\gamma} dq_{3\nu} \times \frac{\rho^4}{N^4}$$

$$- \frac{N(N-1)(N-2)}{2} \sum_{\mu \neq \nu, \gamma} \int f_{\mu\nu}(|q_{1\mu} - q_{2\nu}|) dq_{1\mu} dq_{2\nu} \times \frac{\rho^2}{N^2} \int f_{\gamma\nu}(|q'_{3\gamma} - q'_{3\nu}|) dq'_{3\gamma} dq'_{3\nu} \times \frac{\rho^2}{N^2}$$

By use of indirect correlation function we get:

$$a_{22,32} = \frac{\rho^4}{2N} \sum_{\mu \neq \nu, \gamma} \int f_{\mu\nu}(|q_{1\mu} - q_{2\nu}|) f_{\gamma\nu}(|q_{3\gamma} - q_{3\nu}|) h_\nu(|q_{2\nu} - q_{3\nu}|) dq_{1\mu} dq_{2\nu} dq_{3\gamma} dq_{3\nu}$$

or

$$a_{22,32} = \frac{\rho^3}{2} \sum_{\mu \neq \nu, \gamma} \int f_{\mu\nu}(|q'_{1\mu} - q'_{2\nu}|) f_{\gamma\nu}(|q'_{3\gamma}|) h_\nu(|q'_{2\nu}|) dq'_{1\mu} dq'_{2\nu} dq'_{3\gamma}$$

This is equal to $a_{22,32} = \frac{\rho^3}{2} \sum_{\mu \neq \nu, \gamma} b_{\gamma\nu}(0) \tilde{f}_{\mu\nu}(0) \tilde{h}_\nu(0)$ that vanishes in the thermodynamic limit.

33. In this case we have $i \neq j \neq k$ and $\mu = \gamma$ and $\nu \neq \delta$. Eq. (32) will then become:

$$a_{22,32} = \frac{1}{V^{Nm}} \sum_{\substack{i<j \\ \mu \neq \nu}} \sum_{k \ \delta} \int \frac{e^{-\beta \sum_{\alpha=1}^m U_\alpha}}{Q_0} f_{\mu\nu}(|q_{i\mu} - q_{j\nu}|) f_{\mu\delta}(|q_{k\mu} - q_{k\delta}|) dq_1^N ... dq_m^N$$

$$- \frac{1}{V^{Nm}} \frac{1}{V^{Nm}} \sum_{\substack{i<j \\ \mu \neq \nu}} \sum_{k \ \delta} \int \frac{e^{-\beta \sum_{\alpha=1}^m U_\alpha}}{Q_0} \frac{e^{-\beta \sum_{\alpha=1}^m U_{\alpha'}}}{Q_0} f_{\mu\nu}(|q_{i\mu} - q_{j\nu}|) f_{\mu\delta}(|q'_{k\mu} - q'_{k\delta}|) dq_1^N ... dq_m^N dq_1'^N ... dq_m'^N$$

This is simply equal to:

$$a_{22,33} = \frac{N(N-1)(N-2)}{2} \sum_{\mu \neq \nu, \delta} \int f_{\mu\nu}(|q_{1\mu} - q_{2\nu}|) f_{\mu\delta}(|q_{3\mu} - q_{3\delta}|) g_\mu(|q_{1\mu} - q_{3\mu}|) dq_{1\mu} dq_{2\nu} dq_{3\mu} dq_{3\delta} \times \frac{\rho^4}{N^4}$$

$$- \frac{N(N-1)(N-2)}{2} \sum_{\mu \neq \nu, \delta} \int f_{\mu\nu}(|q_{1\mu} - q_{2\nu}|) dq_{1\mu} dq_{2\nu} \times \frac{\rho^2}{N^2} \int f_{\mu\delta}(|q'_{3\mu} - q'_{3\delta}|) dq'_{3\mu} dq'_{3\delta} \times \frac{\rho^2}{N^2}$$

By use of indirect correlation function we get:

$$a_{22,33} = \frac{\rho^4}{2N} \sum_{\mu \neq \nu, \delta} \int f_{\mu\nu}(|q_{1\mu} - q_{2\nu}|) f_{\mu\delta}(|q_{3\mu} - q_{3\delta}|) h_\mu(|q_{1\mu} - q_{3\mu}|) dq_{1\mu} dq_{2\nu} dq_{3\mu} dq_{3\delta}$$

or

$$a_{22,32} = \frac{\rho^3}{2} \sum_{\mu \neq \nu, \delta} \int f_{\mu\nu}(|q'_{1\mu} - q'_{2\nu}|) f_{\mu\delta}(|q'_{3\delta}|) h_\mu(|q'_{1\mu}|) dq'_{1\mu} dq'_{2\nu} dq'_{3\delta}$$

This is equal to $a_{22,33} = \frac{\rho^3}{2} \sum_{\mu \neq \nu, \delta} b_{\mu\delta}(0) \tilde{f}_{\mu\nu}(0) \tilde{h}_{\mu}(0)$ that vanishes in the thermodynamic limit.

34. In this case we have $i \neq j \neq k$ and $\mu \neq \gamma$ and $\nu \neq \delta$. Eq. (32) will then become:

$$a_{22,34} = \frac{1}{V^{Nm}} \sum_{\substack{i<j \\ \mu \neq \nu}} \sum_{\substack{k \\ \gamma \neq \delta}} \int \frac{e^{-\beta \sum_{\alpha=1}^{m} U_\alpha}}{Q_0} f_{\mu\nu}(|q_{i\mu} - q_{j\nu}|) f_{\mu\delta}(|q_{k\gamma} - q_{k\delta}|) dq_1^N ... dq_m^N$$

$$- \frac{1}{V^{Nm}} \frac{1}{V^{Nm}} \sum_{\substack{i<j \\ \mu \neq \nu}} \sum_{\substack{k \\ \gamma \neq \delta}} \int \frac{e^{-\beta \sum_{\alpha=1}^{m} U_\alpha}}{Q_0} \frac{e^{-\beta \sum_{\alpha=1}^{m} U_{\alpha'}}}{Q_0} f_{\mu\nu}(|q_{i\mu} - q_{j\nu}|) f_{\mu\delta}(|q'_{k\gamma} - q'_{k\delta}|) dq_1^N ... dq_m^N dq_1^{'N} ... dq_m^{'N}$$

This is simply equal to:

$$a_{22,34} = \frac{N^2(N-1)}{2} \sum_{\substack{\mu \neq \nu \\ \gamma \neq \delta}} \int f_{\mu\nu}(|q_{1\mu} - q_{2\nu}|) f_{\gamma\delta}(|q_{3\gamma} - q_{3\delta}|) dq_{1\mu} dq_{2\nu} dq_{3\gamma} dq_{3\delta} \times \frac{\rho^4}{N^4}$$

$$- \frac{N^2(N-1)}{2} \sum_{\substack{\mu \neq \nu \\ \gamma \neq \delta}} \int f_{\mu\nu}(|q_{1\mu} - q_{2\nu}|) dq_{1\mu} dq_{2\nu} \times \frac{\rho^2}{N^2} \int f_{\gamma\delta}(|q'_{3\gamma} - q'_{3\delta}|) dq'_{3\gamma} dq'_{3\delta} \times \frac{\rho^2}{N^2}$$

that is identically zero. So, this group has no contribution at all.

### 4.3. Calculation of $a_{23}$

The explicit form of $a_{23}$ has been given in Eqs. (16), (19), and (21) reads:

$$a_{23} = \frac{1}{V^{Nm}} \sum_{\substack{k \\ \gamma \neq \delta}} \sum_{\substack{k' \\ \gamma' \neq \delta'}} \int \frac{e^{-\beta \sum_{\alpha=1}^{m} U_\alpha}}{Q_0} f_{\gamma\delta}(|q_{k\gamma} - q_{k\delta}|) f_{\gamma'\delta'}(|q_{k'\gamma'} - q_{k'\delta'}|) dq_1^N ... dq_m^N$$

$$- \frac{1}{V^{Nm}} \frac{1}{V^{Nm}} \sum_{\substack{k \\ \gamma \neq \delta}} \sum_{\substack{k' \\ \gamma' \neq \delta'}} \int \frac{e^{-\beta \sum_{\alpha=1}^{m} U_\alpha}}{Q_0} f_{\gamma\delta}(|q_{k\gamma} - q_{k\delta}|) f_{\gamma'\delta'}(|q'_{k'\gamma'} - q'_{k'\delta'}|) dq_1^N ... dq_m^N dq_1^{'N} ... dq_m^{'N}$$

(34)

There are two cases for molecular indexes:

1. $k = k'$ and 2. $k \neq k'$.

to each of which there are four cases from atomic indexes:

1. $\gamma = \gamma'$ and $\delta = \delta'$, 2. $\gamma \neq \gamma'$ and $\delta = \delta'$, 3. $\gamma = \gamma'$ and $\delta \neq \delta'$, and 4. $\gamma \neq \gamma'$ and $\delta \neq \delta'$.

We should calculate $a_{23}$ for all these eight cases.

11. In the first case with $k = k'$, $\gamma = \gamma'$ and $\delta = \delta'$ Eq. (34) becomes:

$$a_{23,11} = \frac{1}{V^{Nm}} \sum_{\substack{k \\ \gamma \neq \delta}} \int \frac{e^{-\beta \sum_{\alpha=1}^{m} U_\alpha}}{Q_0} f_{\gamma\delta}(|q_{k\gamma} - q_{k\delta}|) f_{\gamma\delta}(|q_{k\gamma} - q_{k\delta}|) dq_1^N ... dq_m^N$$

$$- \frac{1}{V^{Nm}} \frac{1}{V^{Nm}} \sum_{\substack{k \\ \gamma \neq \delta}} \int \frac{e^{-\beta \sum_{\alpha=1}^{m} U_\alpha}}{Q_0} f_{\gamma\delta}(|q_{k\gamma} - q_{k\delta}|) f_{\gamma\delta}(|q'_{k\gamma} - q'_{k\delta}|) dq_1^N ... dq_m^N dq_1^{'N} ... dq_m^{'N}$$

This is equal to:

$$a_{23,11} = N \sum_{\gamma \neq \delta} \int f_{\gamma\delta}(|q_{1\gamma} - q_{1\delta}|) f_{\gamma\delta}(|q_{1\gamma} - q_{1\delta}|) dq_{1\gamma} dq_{1\delta} \times \frac{\rho^2}{N^2}$$

$$- N \sum_{\gamma \neq \delta} \int f_{\gamma\delta}(|q_{1\gamma} - q_{1\delta}|) dq_{1\gamma} dq_{1\delta} \times \frac{\rho^2}{N^2} \int f_{\gamma\delta}(|q'_{1\gamma} - q'_{1\delta}|) dq'_{1\gamma} dq'_{1\delta} \times \frac{\rho^2}{N^2}$$

which does not survive in the thermodynamic limit.

12. In this case we have $k = k'$ and $\gamma \neq \gamma'$ and $\delta = \delta'$. Eq. (34) will then become:

$$a_{23,12} = \frac{1}{V^{Nm}} \sum_{\substack{k \\ \gamma \neq \delta}} \sum_{\gamma'} \int \frac{e^{-\beta \sum_{\alpha=1}^{m} U_\alpha}}{Q_0} f_{\gamma\delta}(|q_{k\gamma} - q_{k\delta}|) f_{\gamma'\delta}(|q_{k\gamma'} - q_{k\delta}|) dq_1^N ... dq_m^N$$

$$- \frac{1}{V^{Nm}} \frac{1}{V^{Nm}} \sum_{\substack{k \\ \gamma \neq \delta}} \sum_{\gamma'} \int \frac{e^{-\beta \sum_{\alpha=1}^{m} U_\alpha}}{Q_0} f_{\gamma\delta}(|q_{k\gamma} - q_{k\delta}|) f_{\gamma'\delta}(|q'_{k\gamma'} - q'_{k\delta}|) dq_1^N ... dq_m^N dq_1^{'N} ... dq_m^{'N}$$

This is simply equal to

$$a_{23,12} = N \sum_{\gamma \neq \delta, \gamma'} \int f_{\gamma\delta}(|q_{1\gamma} - q_{1\delta}|) f_{\gamma'\delta}(|q_{1\gamma'} - q_{1\delta}|) dq_{1\gamma} dq_{1\delta} dq_{1\gamma'} \times \frac{\rho^3}{N^3}$$

$$-N \sum_{\gamma \neq \delta, \gamma'} \int f_{\gamma\delta}(|q_{1\gamma} - q_{1\delta}|) dq_{1\gamma} dq_{1\delta} \times \frac{\rho^2}{N^2} \int f_{\gamma'\delta}(|q'_{1\gamma'} - q'_{1\delta}|) dq'_{1\gamma'} dq'_{1\delta} \times \frac{\rho^2}{N^2}$$

which does not survive in the thermodynamic limit.

13. In this case we have $k = k'$ and $\gamma = \gamma'$ and $\delta \neq \delta'$. Eq. (34) will then become:

$$a_{23,13} = \frac{1}{V^{Nm}} \sum_{\substack{k \\ \gamma \neq \delta}} \sum_{\delta'} \int \frac{e^{-\beta \sum_{\alpha=1}^{m} U_\alpha}}{Q_0} f_{\gamma\delta}(|q_{k\gamma} - q_{k\delta}|) f_{\gamma\delta'}(|q_{k\gamma} - q_{k\delta'}|) dq_1^N \ldots dq_m^N$$

$$-\frac{1}{V^{Nm}} \frac{1}{V^{Nm}} \sum_{\substack{k \\ \gamma \neq \delta}} \sum_{\delta'} \int \frac{e^{-\beta \sum_{\alpha=1}^{m} U_\alpha}}{Q_0} f_{\gamma\delta}(|q_{k\gamma} - q_{k\delta}|) f_{\gamma\delta'}(|q'_{k\gamma} - q'_{k\delta'}|) dq_1^N \ldots dq_m^N dq_1'^N \ldots dq_m'^N$$

This is simply equal to

$$a_{23,13} = N \sum_{\gamma \neq \delta, \delta'} \int f_{\gamma\delta}(|q_{1\gamma} - q_{1\delta}|) f_{\gamma\delta'}(|q_{1\gamma} - q_{1\delta'}|) dq_{1\gamma} dq_{1\delta} dq_{1\delta'} \times \frac{\rho^3}{N^3}$$

$$-N \sum_{\gamma \neq \delta, \delta'} \int f_{\gamma\delta}(|q_{1\gamma} - q_{1\delta}|) dq_{1\gamma} dq_{1\delta} \times \frac{\rho^2}{N^2} \int f_{\gamma\delta'}(|q'_{1\gamma} - q'_{1\delta'}|) dq'_{1\gamma} dq'_{1\delta'} \times \frac{\rho^2}{N^2}$$

which does not survive in the thermodynamic limit.

14. In this case we have $k = k'$ and $\gamma \neq \gamma'$ and $\delta \neq \delta'$. Eq. (34) will then become:

$$a_{23,14} = \frac{1}{V^{Nm}} \sum_{\substack{k \\ \gamma \neq \delta \\ \gamma' \neq \delta'}} \int \frac{e^{-\beta \sum_{\alpha=1}^{m} U_\alpha}}{Q_0} f_{\gamma\delta}(|q_{k\gamma} - q_{k\delta}|) f_{\gamma'\delta'}(|q_{k\gamma'} - q_{k\delta'}|) dq_1^N \ldots dq_m^N$$

$$-\frac{1}{V^{Nm}} \frac{1}{V^{Nm}} \sum_{\substack{k \\ \gamma \neq \delta \\ \gamma' \neq \delta'}} \int \frac{e^{-\beta \sum_{\alpha=1}^{m} U_\alpha}}{Q_0} f_{\gamma\delta}(|q_{k\gamma} - q_{k\delta}|) f_{\gamma'\delta'}(|q'_{k\gamma'} - q'_{k\delta'}|) dq_1^N \ldots dq_m^N dq_1'^N \ldots dq_m'^N$$

This is simply equal to

$$a_{23,14} = N \sum_{\substack{\gamma \neq \delta \\ \gamma' \neq \delta'}} \int f_{\gamma\delta}(|q_{1\gamma} - q_{1\delta}|) f_{\gamma'\delta'}(|q_{1\gamma'} - q_{1\delta'}|) dq_{1\gamma} dq_{1\delta} dq_{1\gamma'} dq_{1\delta'} \times \frac{\rho^4}{N^4}$$

$$-N \sum_{\substack{\gamma \neq \delta \\ \gamma' \neq \delta'}} \int f_{\gamma\delta}(|q_{1\gamma} - q_{1\delta}|) dq_{1\gamma} dq_{1\delta} \times \frac{\rho^2}{N^2} \int f_{\gamma'\delta'}(|q'_{1\gamma'} - q'_{1\delta'}|) dq'_{1\gamma'} dq'_{1\delta'} \times \frac{\rho^2}{N^2}$$

which does not survive in the thermodynamic limit.

21. In this case we have $k \neq k'$ and $\gamma = \gamma'$ and $\delta = \delta'$. Eq. (34) will then become:

$$a_{23,21} = \frac{1}{V^{Nm}} \sum_{k} \sum_{\substack{k' \\ \gamma \neq \delta}} \int \frac{e^{-\beta \sum_{\alpha=1}^{m} U_\alpha}}{Q_0} f_{\gamma\delta}(|q_{k\gamma} - q_{k\delta}|) f_{\gamma\delta}(|q_{k'\gamma} - q_{k'\delta}|) dq_1^N ... dq_m^N$$

$$- \frac{1}{V^{Nm}} \frac{1}{V^{Nm}} \sum_{k} \sum_{\substack{k' \\ \gamma \neq \delta}} \int \frac{e^{-\beta \sum_{\alpha=1}^{m} U_\alpha}}{Q_0} f_{\gamma\delta}(|q_{k\gamma} - q_{k\delta}|) f_{\gamma\delta}(|q'_{k'\gamma} - q'_{k'\delta}|) dq_1^N ... dq_m^N dq_1^{'N} ... dq_m^{'N}$$

This is simply equal to

$$a_{23,21} = N^2 \sum_{\gamma \neq \delta} \int f_{\gamma\delta}(|q_{1\gamma} - q_{1\delta}|) f_{\gamma\delta}(|q_{2\gamma} - q_{2\delta}|) g_\gamma(|q_{1\gamma} - q_{2\gamma}|) g_\delta(|q_{1\delta} - q_{2\delta}|) dq_{1\gamma} dq_{1\delta} dq_{2\gamma} dq_{2\delta} \times \frac{\rho^4}{N^4}$$

$$- N^2 \sum_{\gamma \neq \delta} \int f_{\gamma\delta}(|q_{1\gamma} - q_{1\delta}|) dq_{1\gamma} dq_{1\delta} \times \frac{\rho^2}{N^2} \int f_{\gamma\delta}(|q'_{2\gamma} - q'_{2\delta}|) dq'_{2\gamma} dq'_{2\delta} \times \frac{\rho^2}{N^2}$$

Equivalently:

$$a_{23,21} = \frac{\rho^4}{N^2} \sum_{\gamma \neq \delta} \int f_{\gamma\delta}(|q_{1\gamma} - q_{1\delta}|) f_{\gamma\delta}(|q_{2\gamma} - q_{2\delta}|) \left\{ \left[ h_\gamma(|q_{1\gamma} - q_{2\gamma}|) + 1 \right] \left[ h_\delta(|q_{1\delta} - q_{2\delta}|) + 1 \right] - 1 \right\} dq_{1\gamma} dq_{1\delta} dq_{2\gamma} dq_{2\delta}$$

or:

$$a_{23,21} = \frac{\rho^4}{N^2} \sum_{\gamma \neq \delta} \int f_{\gamma\delta}(|q_{1\gamma} - q_{1\delta}|) f_{\gamma\delta}(|q_{2\gamma} - q_{2\delta}|) h_\gamma(|q_{1\gamma} - q_{2\gamma}|) dq_{1\gamma} dq_{1\delta} dq_{2\gamma} dq_{2\delta}$$

$$+ \frac{\rho^4}{N^2} \sum_{\gamma \neq \delta} \int f_{\gamma\delta}(|q_{1\gamma} - q_{1\delta}|) f_{\gamma\delta}(|q_{2\gamma} - q_{2\delta}|) h_\delta(|q_{1\delta} - q_{2\delta}|) dq_{1\gamma} dq_{1\delta} dq_{2\gamma} dq_{2\delta}$$

$$+ \frac{\rho^4}{N^2} \sum_{\gamma \neq \delta} \int f_{\gamma\delta}(|q_{1\gamma} - q_{1\delta}|) f_{\gamma\delta}(|q_{2\gamma} - q_{2\delta}|) h_\gamma(|q_{1\gamma} - q_{2\gamma}|) h_\delta(|q_{1\delta} - q_{2\delta}|) dq_{1\gamma} dq_{1\delta} dq_{2\gamma} dq_{2\delta}$$

which may become simpler:

$$a_{23,21} = \frac{\rho^4 V}{N^2} \sum_{\gamma \neq \delta} \int f_{\gamma\delta}(|q'_{1\delta}|) f_{\gamma\delta}(|q'_{2\gamma} - q'_{2\delta}|) h_\gamma(|q'_{2\gamma}|) dq'_{1\delta} dq'_{2\gamma} dq'_{2\delta}$$

$$+ \frac{\rho^4 V}{N^2} \sum_{\gamma \neq \delta} \int f_{\gamma\delta}(|q'_{1\gamma}|) f_{\gamma\delta}(|q'_{2\gamma} - q'_{2\delta}|) h_\delta(|q'_{2\delta}|) dq'_{1\gamma} dq'_{2\gamma} dq'_{2\delta}$$

$$+ \frac{\rho^4 V}{N^2} \sum_{\gamma \neq \delta} \int f_{\gamma\delta}(|q'_{1\delta}|) h_\delta(|q'_{1\delta} - q'_{2\delta}|) f_{\gamma\delta}(|q'_{2\gamma} - q'_{2\delta}|) h_\gamma(|q'_{2\gamma}|) dq'_{1\delta} dq'_{2\gamma} dq'_{2\delta}$$

This is equal to $a_{23,21} = \frac{\rho^3}{N} \sum_{\gamma \neq \delta} \left( b_{\gamma\delta}^2(0) \left( \tilde{h}_\gamma(0) + \tilde{h}_\delta(0) \right) + \int b_{\gamma\delta}^2(k) \tilde{h}_\gamma(k) \tilde{h}_\delta(k) \frac{d^3 k}{(2\pi)^3} \right)$ which vanishes in the thermodynamic limit.

22. In this case we have $k \neq k'$ and $\gamma \neq \gamma'$ and $\delta = \delta'$. Eq. (34) will then become:

$$a_{23,22} = \frac{1}{V^{Nm}} \sum_{\substack{k \\ \gamma \neq \delta}} \sum_{\substack{k' \\ \gamma'}} \int \frac{e^{-\beta \sum_{\alpha=1}^{m} U_\alpha}}{Q_0} f_{\gamma\delta}(|q_{k\gamma} - q_{k\delta}|) f_{\gamma'\delta}(|q_{k'\gamma'} - q_{k'\delta}|) dq_1^N ... dq_m^N$$

$$- \frac{1}{V^{Nm}} \frac{1}{V^{Nm}} \sum_{\substack{k \\ \gamma \neq \delta}} \sum_{\substack{k' \\ \gamma'}} \int \frac{e^{-\beta \sum_{\alpha=1}^{m} U_\alpha}}{Q_0} f_{\gamma\delta}(|q_{k\gamma} - q_{k\delta}|) f_{\gamma'\delta}(|q'_{k'\gamma'} - q'_{k'\delta}|) dq_1^N ... dq_m^N dq_1^{'N} ... dq_m^{'N}$$

This is simply equal to

$$a_{23,22} = N^2 \sum_{\gamma \neq \delta, \gamma'} \int f_{\gamma\delta}(|q_{1\gamma} - q_{1\delta}|) f_{\gamma'\delta}(|q_{2\gamma'} - q_{2\delta}|) g_\delta(|q_{1\delta} - q_{2\delta}|) dq_{1\gamma} dq_{1\delta} dq_{2\gamma'} dq_{2\delta} \times \frac{\rho^4}{N^4}$$

$$- N^2 \sum_{\gamma \neq \delta, \gamma'} \int f_{\gamma\delta}(|q_{1\gamma} - q_{1\delta}|) dq_{1\gamma} dq_{1\delta} \times \frac{\rho^2}{N^2} \int f_{\gamma'\delta}(|q'_{2\gamma'} - q'_{2\delta}|) dq'_{2\gamma'} dq'_{2\delta} \times \frac{\rho^2}{N^2}$$

Equivalently:

$$a_{23,22} = \frac{\rho^4}{N^2} \sum_{\gamma \neq \delta, \gamma'} \int f_{\gamma\delta}(|q_{1\gamma} - q_{1\delta}|) f_{\gamma'\delta}(|q_{2\gamma'} - q_{2\delta}|) h_\delta(|q_{1\delta} - q_{2\delta}|) dq_{1\gamma} dq_{1\delta} dq_{2\gamma'} dq_{2\delta}$$

or:

$$a_{23,22} = \frac{\rho^3}{N} \sum_{\gamma \neq \delta, \gamma'} \int f_{\gamma\delta}(|q'_{1\delta}|) f_{\gamma'\delta}(|q'_{2\gamma'} - q'_{2\delta}|) h_\delta(|q'_{1\delta} - q'_{2\delta}|) dq'_{1\delta} dq'_{2\gamma'} dq'_{2\delta}$$

and hence $a_{23,22} = \frac{\rho^3}{N} \sum_{\gamma \neq \delta, \gamma'} b_{\gamma\delta}(0) b_{\gamma'\delta}(0) \tilde{h}_\delta(0)$ that vanishes in the thermodynamic limit.

23. In this case we have $k \neq k'$ and $\gamma = \gamma'$ and $\delta \neq \delta'$. Eq. (34) will then become:

$$a_{23,23} = \frac{1}{V^{Nm}} \sum_{\substack{k \\ \gamma \neq \delta}} \sum_{\substack{k' \\ \delta'}} \int \frac{e^{-\beta \sum_{\alpha=1}^{m} U_\alpha}}{Q_0} f_{\gamma\delta}(|q_{k\gamma} - q_{k\delta}|) f_{\gamma\delta'}(|q_{k'\gamma} - q_{k'\delta'}|) dq_1^N ... dq_m^N$$

$$- \frac{1}{V^{Nm}} \frac{1}{V^{Nm}} \sum_{\substack{k \\ \gamma \neq \delta}} \sum_{\substack{k' \\ \delta'}} \int \frac{e^{-\beta \sum_{\alpha=1}^{m} U_\alpha}}{Q_0} f_{\gamma\delta}(|q_{k\gamma} - q_{k\delta}|) f_{\gamma\delta'}(|q_{k'\gamma} - q_{k'\delta'}|) dq_1^N ... dq_m^N dq_1^{'N} ... dq_m^{'N}$$

This is simply equal to

$$a_{23,23} = N^2 \sum_{\gamma \neq \delta, \delta'} \int f_{\gamma\delta}(|q_{1\gamma} - q_{1\delta}|) f_{\gamma\delta'}(|q_{2\gamma} - q_{2\delta'}|) g_\gamma(|q_{1\gamma} - q_{2\gamma}|) dq_{1\gamma} dq_{1\delta} dq_{2\gamma} dq_{2\delta'} \times \frac{\rho^4}{N^4}$$

$$- N^2 \sum_{\gamma \neq \delta, \delta'} \int f_{\gamma\delta}(|q_{1\gamma} - q_{1\delta}|) dq_{1\gamma} dq_{1\delta} \times \frac{\rho^2}{N^2} \int f_{\gamma\delta'}(|q'_{2\gamma} - q'_{2\delta'}|) dq'_{2\gamma} dq'_{2\delta'} \times \frac{\rho^2}{N^2}$$

Equivalently:

$$a_{23,23} = \frac{\rho^4}{N^2} \sum_{\gamma \neq \delta, \delta'} \int f_{\gamma\delta}(|q_{1\gamma} - q_{1\delta}|) f_{\gamma\delta'}(|q_{2\gamma} - q_{2\delta'}|) h_\gamma(|q_{1\gamma} - q_{2\gamma}|) dq_{1\gamma} dq_{1\delta} dq_{2\gamma} dq_{2\delta'}$$

or

$$a_{23,23} = \frac{\rho^3}{N} \sum_{\gamma \neq \delta, \delta'} \int f_{\gamma\delta}(|q'_{1\delta}|) f_{\gamma\delta'}(|q'_{2\gamma} - q'_{2\delta'}|) h_\gamma(|q'_{2\gamma}|) dq'_{1\delta} dq'_{2\gamma} dq'_{2\delta'}$$

and hence $a_{23,23} = \frac{\rho^3}{N} \sum_{\gamma \neq \delta, \delta'} b_{\gamma\delta}(0) b_{\gamma\delta'}(0) \tilde{h}_\gamma(0)$ that vanishes in the thermodynamic limit.

24. In this case we have $k \neq k'$ and $\gamma \neq \gamma'$ and $\delta \neq \delta'$. Eq. (34) will then become:

$$a_{23,24} = \frac{1}{V^{Nm}} \sum_{\substack{k \\ \gamma \neq \delta}} \sum_{\substack{k' \\ \gamma' \neq \delta'}} \int \frac{e^{-\beta \sum_{\alpha=1}^{m} U_\alpha}}{Q_0} f_{\gamma\delta}(|q_{k\gamma} - q_{k\delta}|) f_{\gamma'\delta'}(|q_{k'\gamma'} - q_{k'\delta'}|) dq_1^N \ldots dq_m^N$$

$$- \frac{1}{V^{Nm}} \frac{1}{V^{Nm}} \sum_{\substack{k \\ \gamma \neq \delta}} \sum_{\substack{k' \\ \gamma' \neq \delta'}} \int \frac{e^{-\beta \sum_{\alpha=1}^{m} U_\alpha}}{Q_0} f_{\gamma\delta}(|q_{k\gamma} - q_{k\delta}|) f_{\gamma'\delta'}(|q'_{k'\gamma'} - q'_{k'\delta'}|) dq_1^N \ldots dq_m^N dq_1'^N \ldots dq_m'^N$$

This is simply equal to

$$a_{23,24} = N^2 \sum_{\substack{\gamma \neq \delta \\ \gamma' \neq \delta'}} \int f_{\gamma\delta}(|q_{1\gamma} - q_{1\delta}|) f_{\gamma'\delta'}(|q_{2\gamma'} - q_{2\delta'}|) dq_{1\gamma} dq_{1\delta} dq_{2\gamma'} dq_{2\delta'} \times \frac{\rho^4}{N^4}$$

$$- N^2 \sum_{\substack{\gamma \neq \delta \\ \gamma' \neq \delta'}} \int f_{\gamma\delta}(|q_{1\gamma} - q_{1\delta}|) dq_{1\gamma} dq_{1\delta} \times \frac{\rho^2}{N^2} \int f_{\gamma'\delta'}(|q'_{2\gamma'} - q'_{2\delta'}|) dq'_{2\gamma'} dq'_{2\delta'} \times \frac{\rho^2}{N^2}$$

that is identically zero.

We thus calculated all contributions due to second perturbation term. As can be observed intra-atomic potentials do not have any contribution to the excess free energy of a molecular fluid and hence all its excess thermodynamic properties over ideal parts are due to van der Waals forces in this theory. Gathering all surviving terms, the final expression for the reduced excess free energy of a molecular fluid takes the following form:

$$a = \sum_{\mu} a_{0\mu} - \frac{\rho}{2} \sum_{\mu \neq \nu} \tilde{f}_{\mu\nu}(0) - \frac{\rho}{4} \sum_{\mu \neq \nu} \widetilde{f_{\mu\nu}^2}(0)$$

$$- \frac{\rho^2}{4} \sum_{\mu \neq \nu} \int \tilde{f}_{\mu\nu}^2(k) \left( \tilde{h}_\mu(k) + \tilde{h}_\nu(k) \right) \frac{d^3k}{(2\pi)^3} - \frac{\rho^3}{8} \sum_{\mu \neq \nu} \tilde{f}_{\mu\nu}(0) \left( \sum_\lambda \tilde{f}_{\mu\lambda}(0) \tilde{h}_\mu(0) + \sum_\lambda \tilde{f}_{\lambda\nu}(0) \tilde{h}_\nu(0) \right) \quad (35)$$

$$- \frac{\rho^3}{8} \sum_{\mu \neq \nu} \left( \tilde{f}_{\mu\nu}^2(0) \tilde{h}_\mu(0) + \tilde{f}_{\mu\nu}^2(0) \tilde{h}_\nu(0) + \int \tilde{f}_{\mu\nu}^2(k) \tilde{h}_\mu(k) \tilde{h}_\nu(k) \frac{d^3k}{(2\pi)^3} \right)$$

The equation may be transformed into much simpler and symmetrical form if we assume that $f_{\mu\mu}(|q_{1\mu} - q_{2\mu}|) = 0$ in perturbation terms. Note that we have already accounted for these (direct) potentials in reference system. Here we make this assumption in order to symmetrize and simplify the perturbation terms, as well as removing the restrictions on summations. That is, by use of this convention all indices run over all atoms (from 1 to $m$). We thus find (after some simple algebra):

$$a = \sum_{\mu=1}^{m} a_{0\mu} - \frac{\rho}{2} \sum_{\mu,\nu=1}^{m} \left( \tilde{f}_{\mu\nu}(0) + \widetilde{f_{\mu\nu}^2}(0)/2 \right)$$

$$- \frac{\rho^3}{4} \sum_{\mu=1}^{m} \tilde{h}_\mu(0) \left( \sum_{\nu=1}^{m} \tilde{f}_{\mu\nu}(0) \right)^2 - \frac{\rho^2}{2} \sum_{\mu,\nu=1}^{m} \int \tilde{f}_{\mu\nu}^2(k) \left( \tilde{h}_\mu(k) + \rho \tilde{h}_\mu(k) \tilde{h}_\nu(k)/4 \right) \frac{d^3k}{(2\pi)^3} \quad (36)$$

This is our sought-for expression for the excess free energy of molecular fluids which can be used to deliver their thermodynamics based on our second order perturbation theory. As said before, there is no contribution from intra-potentials at all. This is, of course, fortunate for this perturbation theory because intra-potentials are much stronger than inter-potentials, and it is not generally safe for a perturbation theory to regard stronger terms as perturbative ones. So we can conclude that, at least based on second order perturbation theory, intra-potentials has nothing to do with the thermodynamics of complex fluids other than ideal contributions. So, based on this theory, molecular fluids without van der Waals forces should be regarded like molecular ideal gases. In other words, like Born-Oppenheimer separation in molecular quantum mechanics, we can analogously divide intra-potentials and inter-potentials contributions and study them separately. Thus, intra-potentials contributions can be accounted for without introducing inter-potentials, i.e., like an ideal gas of non-interacting molecules, and then should be added to the contribution due to van der Waals forces. Furthermore, this also proves, up to the second order perturbation theory, that the molecules made up of atoms by intra-potentials preserve their identities and shapes whether in their ideal gas states or in their condensed states. In fact, the shape and geometry of the molecules are all set and fixed by intra-atomic potentials, but unfortunately do not enter in the theory. On the other hand, we know that both geometry and intra-atomic potentials of the molecules play important role in their thermodynamic properties especially equation of state. We can thus conclude that one should complement the theory by inclusion of these effects in some unknown way. This means that the theory presented here can only been suitable for fluids with small rigid molecules at low densities, because these effects have much less contributions for such molecules. When excess properties due to van der Waals forces are to be computed, it suffices to employ Eq. (36). To complete the theory at this stage (in the lack of information on shape and intra-atomic potentials) one should add ideal contributions to excess properties computable by this perturbation theory. Excess thermodynamic properties such as internal energy, compressibility factor (or pressure), and also chemical potential of molecular fluids are easily obtainable by use of Eq. (36) presented above. We will use $Z = \rho \left( \frac{\partial a}{\partial \rho} \right)_T$ and $P = Z\rho kT$ to obtain compressibility factor and pressure and $U = \left( \frac{\partial a}{\partial \beta} \right)_\rho$ to obtain excess internal energy. In the next section we will employ these formulas to obtain the thermodynamic properties of two systems; hard sphere chain and carbon dioxide molecular fluids together with some details of calculations.

### III. Applications to Hard Sphere Chain and Carbon Dioxide Molecular Fluids and Discussions

In this section we briefly examine the present theory and apply it on two molecular fluids with known thermodynamic properties; hard sphere chain fluid as a well-known model and carbon dioxide as a real system. Thermodynamics of hard sphere chain fluid has been thoroughly investigated both theoretically and by simulations [1-2, 20]. It is a fluid composed of hard spheres as monomers glued together to form identical chains. It is assumed that all chains have the same number of monomers (or hard spheres). Its accurate equation of state reads (called Wertheim TPT1 equation):

$$Z = m\frac{\eta^3 - \eta^2 - \eta - 1}{(\eta - 1)^3} - (m-1)\frac{1 - \eta - \eta^2/2}{(1-\eta)(1-\eta/2)} \qquad (37)$$

where $m$ represents the number of monomers in a chain and $\eta\left(=\frac{\pi}{6}m\rho\sigma^3\right)$ the volume fraction of the hard sphere monomers each with diameter $\sigma$. We will as usual set $\sigma = 1$. Now let us apply the theory presented here that is Eq. (36) to this model. Based on the theory we have $m^2$ hard sphere potentials from which $m$ direct ones are considered as the reference system and the remaining ($m^2-m$) ones are considered as the perturbation potentials. It is simple to calculate all needed ingredients; $\tilde{f}_{\mu\nu}(0) = -\frac{4}{3}\pi$, $\widetilde{f_{\mu\nu}^2}(0) = \frac{4}{3}\pi$, and $\tilde{h}_\mu(0) = \frac{S(0)-1}{\rho}$ where $S(0) = k_B T\left(\frac{\partial\rho}{\partial p}\right)_{T,V}$. For hard spheres (based on Carnahan-Starling [23] equation of state) it is given by $S(0) = \dfrac{(1-\frac{\pi}{6}\rho)^4}{\frac{\pi^4}{1296}\rho^4 - \frac{\pi^3}{54}\rho^3 + \frac{\pi^2}{9}\rho^2 + \frac{2\pi}{3}\rho + 1}$. Furthermore, $\tilde{f}_{\mu\nu}^2(k)$ can be best approximated by delta function for hard sphere potentials, i.e. $\tilde{f}_{\mu\nu}^2(k) \approx \delta(k)$. The final result will be:

$$a = \underbrace{a_{IG}}_{\substack{ideal\ gas\\contribution}} + \underbrace{m a_{HS}}_{\substack{reference\ system\\contribution}} \underbrace{-\left(\frac{m(m-1)}{2}(-\frac{4}{3}\pi)\rho\right)}_{first\ perturbation\ contribution} - \underbrace{\left(\frac{m(m-1)}{4}(\frac{4}{3}\pi)\rho + \frac{m(m-1)^2}{4}(S(0)-1)(-\frac{4}{3}\pi)^2\rho^2 + \frac{m(m-1)}{2}\left((S(0)-1)+(S(0)-1)^2/4\right)\rho\right)}_{second\ perturbation\ contribution} \qquad (38)$$

where we have added the ideal gas contribution in order to complete the theory (although it is not sufficient because of the lack of the intra-molecular potentials contributions). We will then be able to compute the compressibility factor of hard sphere chain fluid and compare the results to those due to Eq.

(37) and computer simulation data [24, 25]. The results of such computations for $m = 4, 8, 16, 51,$ and 201 are given in **Figs**. **I** - **V**. As the figures show the present theory performs well in low density region for all cases and fairly good up to intermediate densities for short-length chains. As the chain grows the agreement is deteriorated. This can be attributed to the inability of the present theory to account for intra-potential contributions. In all cases, of course, the theory works well for all chains at low densities. If one could find a way to include the shape and geometry of the molecule or intra-molecular potentials in the theory, one then could calculate the thermodynamics of molecular liquids at all densities and temperatures confidently. Clearly the theory works well when intra-molecular potentials have less role in thermodynamic properties. This happens naturally at low densities, high temperatures, and small size of molecules. Let us make these conditions transparent by working on a molecular fluid with small molecules (and the simplest geometry) at low densities and high temperatures, i.e. carbon dioxide ($CO_2$). The selected force field is that of Zhang and Duan [26]. The adopted potential model between every two atoms are consisted of a Lennard-Jones and a Coulomb ones. The parameters are $\varepsilon_{C-C}/k_B = 28.845 \ K$, $\sigma_{C-C} = 2.7918$ Å, $\varepsilon_{O-O}/k_B = 82.656 \ K$, $\sigma_{O-O} = 3.00$ Å, and the partial charges on oxygen and carbon are assumed to be 0.5888 $e$ and -0.2944 $e$, respectively. For unlike atoms the Lorentz-Berthelot combining rule is assumed. For carbon dioxide fluid we need to have the structure and thermodynamics of two simple atomic fluids composed of atomic oxygens and carbons. Both integral equation and perturbation theories can be invoked for this purpose. Because of higher accuracy of perturbation theories in giving both aspects we opt to employ the simplest, fastest, and accurate one, the Optimized Random Phase Approximation (ORPA). We thus calculated the structure and thermodynamics of both oxygen and carbon atomic fluids by use of ORPA. During calculations we selected $\varepsilon_{C-C}$ and $\sigma_{C-C}$ to reduce the temperature and the density (we could have used those of oxygen with essentially the same results). By use of them all three potentials at first are reduced and then employed in computations. Second step is to perform the calculations of these simple liquids based on ORPA. The details of such standard numerical calculations have been given for example in refs. 14-15 and avoided here. Of course, one should be careful of reduced densities used for this calculations. Since every potentials has its own hard core diameter, calculated by Barker-Henderson criterion [27], which is used in all subsequent calculations in ORPA or every perturbation theory, one should alter the reduced density on this basis. We can then calculated simply both the structure and thermodynamics of the reference simple liquids. They are now our needed inputs to be plugged into Eq. (36) from which thermodynamic properties can easily be computed. The final step is to account for ideal gas contributions that is adding $Z^{IG}=1$ to the excess compressibility factor and $U^{IG}=\frac{13}{2}RT$ to the excess internal energy. Note that the rotational and vibrational contributions are also accounted for in addition to translational one. Because of these contributions that are due to the special geometry of the molecule even in ideal state, we can expect that the calculated internal energy should be much more accurate than the compressibility factor and hence the pressure. If we could in some way account for the effects of molecular geometry into the scheme even without van der Waals forces, then the present theory could have captured and given essentially the thermodynamics of molecular fluids accurately. But, unfortunately, we do not have such a theory. The results of such calculations together with experimental values at the corresponding states from NIST [28] are brought in **Tables 1-5**. As expected, there is fairly good agreement for pressure and to some extent for internal energy at lowest studied density presented in **table. 1**. When the density is increased the agreement starts to deteriorate. The best agreements happen in the gaseous (both vapor and supercritical) states whereas the worst ones happen at the vapor–liquid coexisting regions. At high densities and low temperatures where the system falls into the liquid states, the internal potentials and the geometry of molecules get important such that their roles cannot be ignored as in the present theory. So the present perturbation theory is appropriate to present the gaseous thermodynamic properties of molecular fluids.

**IV. Conclusion**

A new perturbation theory for molecular liquids has been developed where a new form of splitting the site-site potential functions between molecules is introduced. By which a set of atomic fluids can be considered as the reference system with known structure and thermodynamics. The perturbative part of the potential function has then be expanded up to two terms. The excess Helmholtz free energy of the system has then been obtained for three computable contributions. The derivation has shown that the excess Helmholtz free energy cannot incorporate the intra-atomic potentials, all contributions come merely from inter-atomic potentials. Then it has been applied to compute the thermodynamics of two systems; hard sphere chain and carbon dioxide molecular fluids. The results compared with the computer simulation data show that the theory works well at low densities and gaseous states.

## V. Acknowledgement

I would like to thank the Islamic Azad University, Amol Branch for generous financial support of the research project.

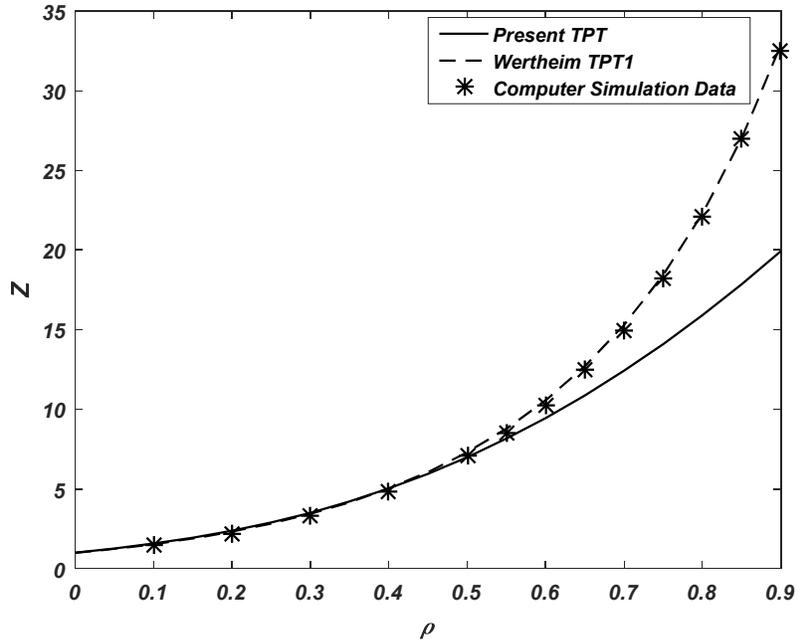

**Fig. 1**: The compressibility factor $Z\left(=\dfrac{PV}{NkT}\right)$ of hard sphere chain with $m = 4$ predicted by present theory, thermodynamic perturbation theory (TPT1) of Wertheim at various densities compared with the experimental data taken from ref. 24.

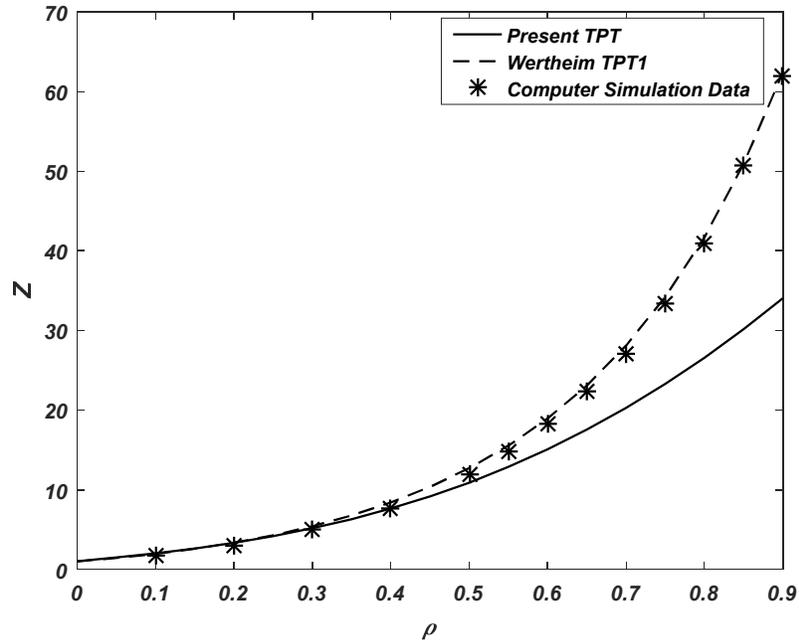

**Fig**. 2: The same as **Fig**. 1 but for $m = 8$.

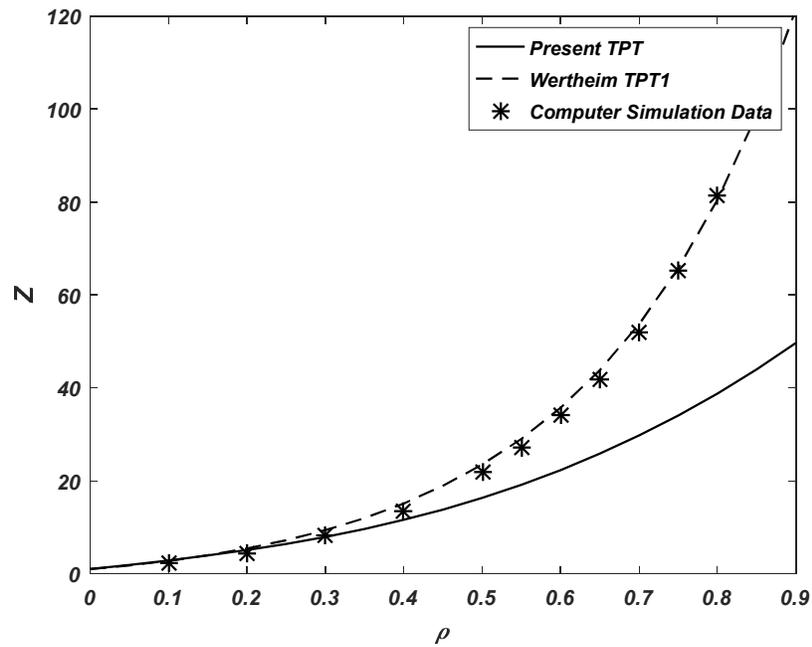

**Fig**. 3: The same as **Fig**. 1 but for $m = 16$.

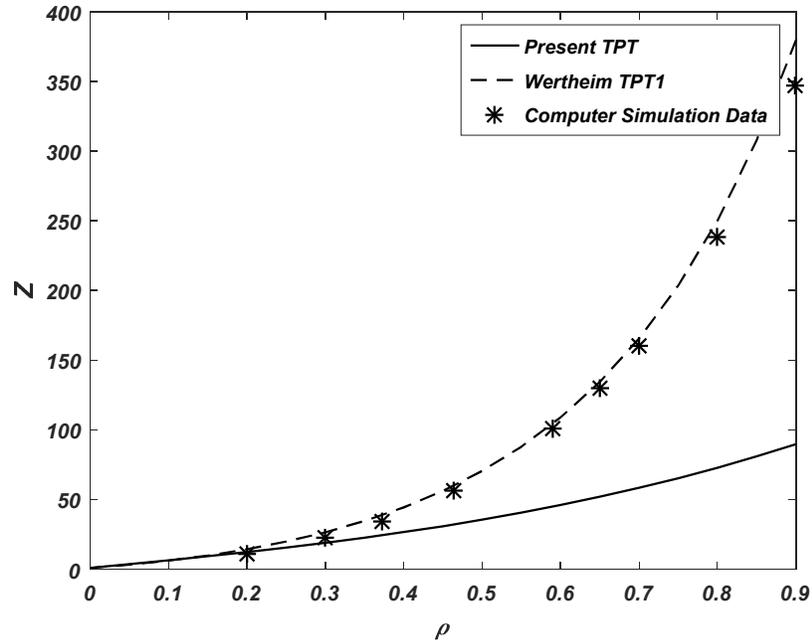

**Fig**. 4: The same as **Fig**. 1 but for *m* = 51 and experimental data are taken from ref. 25.

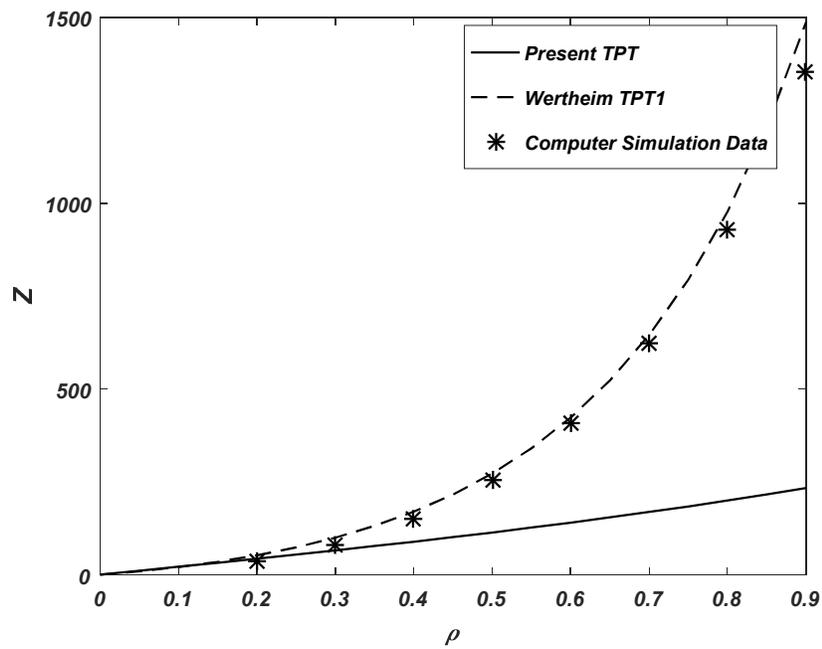

**Fig**. 5: The same as **Fig**. 1 but for *m* = 201 and experimental data are taken from ref. 25.

.

**Table. 1:** Thermodynamic properties of carbon dioxide predicted by present theory at constant density $\rho = 0.05$ gr/ml and various temperatures compared with the experimental data taken from NIST [28].

| states at $\rho = 0.05$ gr/ml | $T$(k) | $P$(bar) Present work | experimental | $U$(kJ/mol) Present work | experimental |
|---|---|---|---|---|---|
| vapor | 250 | 20.523 | 18.157 | 13.446 | 17.551 |
| vapor | 300 | 25.813 | 24.625 | 16.151 | 19.101 |
| vapor | 400 | 36.285 | 35.424 | 21.559 | 22.412 |
| vapor | 600 | 57.006 | 56.011 | 32.372 | 29.782 |
| supercritical | 800 | 77.593 | 76.267 | 43.183 | 38.068 |
| supercritical | 950 | 92.95 | 91.408 | 51.291 | 44.746 |

**Table. 2:** The same as table. 1 but at $\rho = 0.1$ gr/ml.

| states at $\rho = 0.1$ gr/ml | $T$(k) | $P$(bar) Present work | experimental | $U$(kJ/mol) Present work | experimental |
|---|---|---|---|---|---|
| Vapor and liquid | 250 | 35.636 | 18.157 | 13.382 | 11.456 |
| vapor | 300 | 47.443 | 42.497 | 16.09 | 18.34 |
| vapor | 400 | 70.65 | 66.576 | 21.502 | 21.838 |
| supercritical | 600 | 116.24 | 111.15 | 32.319 | 29.331 |
| supercritical | 800 | 161.34 | 154.5 | 43.134 | 37.669 |
| supercritical | 950 | 194.81 | 186.75 | 51.244 | 44.373 |

**Table. 3:** The same as table. 1 but at $\rho = 0.2$ gr/ml.

| states at $\rho = 0.2$ gr/ml | $T$(k) | $P$(bar) Present work | experimental | $U$(kJ/mol) Present work | experimental |
|---|---|---|---|---|---|
| Vapor and liquid | 250 | 54.097 | 18.157 | 13.257 | 8.701 |
| vapor | 300 | 82.948 | 62.329 | 15.972 | 16.887 |
| supercritical | 400 | 139.25 | 119.4 | 21.391 | 20.749 |
| supercritical | 600 | 249.49 | 222.18 | 32.217 | 28.453 |
| supercritical | 800 | 358.28 | 320.92 | 43.037 | 36.887 |
| supercritical | 950 | 438.45 | 393.93 | 51.152 | 43.641 |

**Table. 4:** The same as table. 1 but at $\rho = 0.3$ gr/ml.

| states at $\rho = 0.3$ gr/ml | $T$(k) | $P$(bar) Present work | experimental | $U$(kJ/mol) Present work | experimental |
|---|---|---|---|---|---|
| Vapor and liquid | 250 | 63.591 | 18.157 | 13.139 | 7.783 |
| Vapor and liquid | 300 | 114.83 | 67.37 | 15.861 | 15.292 |
| supercritical | 400 | 214.6 | 165 | 21.286 | 19.732 |
| supercritical | 600 | 411.94 | 341 | 32.119 | 27.596 |
| supercritical | 800 | 608.06 | 509.68 | 42.944 | 36.116 |
| supercritical | 950 | 752.16 | 633.85 | 51.064 | 42.921 |

**Table. 5:** The same as table. 1 but at $\rho = 0.5$ gr/ml.

| states at $\rho = 0.5$ gr/ml | $T$(k) | $P$(bar) Present work | experimental | $U$(kJ/mol) Present work | experimental |
|---|---|---|---|---|---|
| Vapor and liquid | 250 | 81.619 | 18.157 | 12.924 | 7.049 |
| Vapor and liquid | 300 | 189.28 | 67.37 | 15.673 | 12.974 |
| supercritical | 400 | 399.2 | 261.8 | 21.113 | 17.837 |
| supercritical | 600 | 839.56 | 646.62 | 31.949 | 25.894 |
| supercritical | 800 | 1294.8 | 1016.2 | 42.779 | 34.577 |
| supercritical | 950 | 1631.7 | 1285.8 | 50.906 | 41.495 |